\documentclass[iop,revtex4]{emulateapj}
\usepackage{natbib,textcomp}
\usepackage{lscape}

\def\xmm{{\it XMM-Newton}}
\def\akari{{\it AKARI}}

\def\eps{{erg s$^{-1}$}}
\def\pcm{{cm$^{-2}$}}
\def\NH{$N_{\rm H}$}

\slugcomment{Accepted for publication in the {\it Astrophysical Journal}}

\shorttitle{X-ray and IR selection of Obscured AGNs}

\shortauthors{Terashima et al.}

\begin{document}

\title{A New Sample of Obscured AGNs Selected from the XMM-Newton and AKARI Surveys}

\author{Yuichi Terashima\altaffilmark{1}, 
Yoshitaka Hirata\altaffilmark{1},
Hisamitsu Awaki\altaffilmark{1},
Shinki Oyabu\altaffilmark{2},
Poshak Gandhi\altaffilmark{3},\\
Yoshiki Toba\altaffilmark{4},
and Hideo Matsuhara\altaffilmark{5}
}
\altaffiltext{1}{Department of Physics, Ehime University, Bunkyo-cho, Matsuyama, Ehime 790-8577, Japan}

\altaffiltext{2}{Department of Physics, Nagoya University, Furo-cho, Chikusa-ku, Nagoya, Aichi 464-8602, Japan}

\altaffiltext{3}{School of Physics \& Astronomy, University of Southampton, Highfield, Southampton, SO17 1BJ, UK}

\altaffiltext{4}{Research Center for Space and Cosmic Evolution, Ehime University, Bunkyo-cho, Matsuyama, Ehime 790-8577, Japan}

\altaffiltext{5}{Institute of Space and Astronautical Science, 3-1-1 Yoshinodai, Chuo-ku, Sagamihara, Kanagawa 252-5210, Japan}

\begin{abstract}

We report a new sample of obscured active galactic nuclei (AGNs) selected from the \xmm~ serendipitous source 
and \akari~point-source catalogs. We match X-ray sources with infrared (18 and 90 $\mu$m) sources located at $|b|>10^\circ$
to create a sample consisting of 173 objects. Their optical classifications and absorption column densities 
measured by X-ray spectra are compiled and study efficient selection criteria to find obscured AGNs.
We apply the criteria (1) X-ray hardness ratio defined by using the $2-4.5$ keV and $4.5-12$ keV bands $>-0.1$
and (2) EPIC-PN count rate (CR) in the $0.2-12$ keV  to infrared flux ratio CR/$F_{90}<0.1$ or CR/$F_{18}<1$,
where  $F_{18}$ and $F_{90}$ are infrared fluxes at 18 and 90 $\mu$m in Jy, respectively, to search for
obscured AGNs.
X-ray spectra of 48 candidates, for which no X-ray results have been published,
are analyzed and X-ray evidence for the presence of obscured AGNs such as a convex shape X-ray spectrum
indicative of absorption of \NH~$\sim10^{22-24}$ \pcm, a very flat continuum, or a strong Fe-K emission line
with an equivalent width of $>700$ eV is found in 26 objects. 
Six among them are classified as Compton-thick AGNs, and four are represented by either Compton-thin or Compton-thick
spectral models.
The success rate of finding obscured AGNs combining our analysis and the literature is 92\% if the 18 $\mu$m condition is used.
Of the 26 objects, 4 are optically classified as an \ion{H}{2} nucleus and are new  ``elusive AGNs'' in which
star formation activity likely overwhelms AGN emission in the optical and infrared bands.

\end{abstract}


\section{INTRODUCTION}

Multiwavelength observations have been finding various populations of active galactic nuclei (AGN). 
The population of obscured 
AGNs, which constitute a large fraction of AGNs (e.g., Fabian 2004, Gilli et al. 2007), among various classes,
are believed to be important in various aspects including the origin of the Cosmic X-ray 
background (XRB; e.g., Gilli et al. 2007), connection between obscuring matter and star formation 
activity in the host galaxies (e.g., Wada \& Norman 2002), and evolutional paths of AGNs (e.g., Sanders et al., 1988, Hopkins et al. 2006, 
Alexander \& Hickox 2012).
Modern large area surveys  at various wavebands are indeed utilized to find a large number of hidden AGNs and to elucidate
their nature.

Obscured AGNs are found by X-ray emission transmitted through obscuring matter, optical line emission from 
extended and ionized regions, infrared emission from dust heated by central AGNs, and so on. 
Hard X-ray surveys are one effective way to find obscured AGNs because the photoelectric
cross section decreases as X-ray energy increases, and transmitted X-rays can be observed.
Indeed, more than a dozen of Compton-thick AGNs, which are absorbed by a hydrogen column density 
greater than $1.5\times10^{24}$ cm$^{-2}$, show transmitted hard X-rays above 10 keV 
(Comastri 2004 and references therein; Burlon et al. 2011).
The all-sky and sensitive hard X-ray surveys conducted by the {\it Swift} Burst Alert Telescope (BAT) and {\it INTEGRAL} provide an
unprecedented opportunity to search for the heavily absorbed population. While obscured AGNs have been
found in these surveys, the number of Compton-thick sources are not as many as expected from other 
studies using, e.g.,  optical emission-line-selected samples (Gilli et al. 2007) or prediction 
from the synthesis models of the XRB (Burlon et al. 2011). This bias is due to the attenuation of X-rays even at
energies above 10 keV for Compton-thick cases 
(Wilman \& Fabian 1999, Ikeda et al. 2009, Murphy \& Yaqoob 2009, Brightman \& Nandra 2011a).
Burlon et al. (2011) estimated a correction factor of about four for Compton-thick fraction in their {\it Swift}/BAT sample.
Thus, hard X-rays are still biased against heavily absorbed AGNs.

Infrared emission from warm dust heated by the central source is also employed to find obscured AGNs. 
Mid-infrared (MIR) emission indeed traces the power of the AGNs regardless of whether they are type 1 or 2, 
and the attenuation of MIR in
type 2 AGNs is not large (Horst et al. 2008, Gandhi et al. 2009, Ichikawa et al. 2012, Matsuta et al. 2012).
A weakness of utilizing infrared emission, however, is that emission from dust heated by stars cannot be
separated if spatial resolution is not sufficiently high, and therefore the infrared-selected sample contains
non-negligible fraction of non-AGN galaxies. For example, 13 and 18 out of 126 galaxies
selected at 12 $\mu$m
are optically classified as \ion{H}{2} and \ion{H}{2}/AGN composite galaxies, respectively
 (Brightman \& Nandra 2011b).

The combination of X-ray and infrared selection is a useful way to select a heavily 
obscured AGN population and is applied to deep (Fiore et al. 2008, 2009)
or wide surveys (Mateos et al. 2012, Severgnini et al. 2012, Rovilos et al. 2014) to overcome the biases in the selections using
only hard X-ray or infrared emission.
The X-ray to infrared flux ratios, X-ray hardness, infrared colors, and so on are utilized in these selections.
Among the techniques employed, we extend the method used by Severgnini et al. (2012).
They used 25 $\mu$m fluxes ($F_{25})$ measured in the {\it IRAS} Point Source Catalog (PSC)
and X-ray data taken from the {\xmm} serendipitous source catalog (2XMM catalog; Watson et al. 2009)
and made a diagnostic plot of X-ray hardness ratio (HR4) and X-ray to infrared flux ratio ($F$(2$-$12 keV)/$\nu_{25}F_{25}$),
where HR4 is defined by using X-ray count rates (CRs) in 2--4.5 keV CR(2--4.5 keV) and in 4.5--12 keV CR(4.5--12 keV) as 
\[
{\rm HR4} = \frac{{\rm CR(4.5-12~keV}) - {\rm CR(2.0-4.5~keV)}}{{\rm CR(4.5-12~keV)} + {\rm CR(2.0-4.5~keV)}}.
\]
They defined the region for candidates of Compton-thick AGNs as $F{\rm (2-12 keV})/\nu_{25}F_{25} < 0.02$ 
and HR4$>-0.2$, and built a sample consisting of 43 candidates. For absorbed sources, the X-ray to infrared ratio becomes small
since X-rays below 12 keV are attenuated by photoelectric absorption. Absorbed sources show flatter X-ray spectra and
therefore larger values of hardness ratios are expected. Thus, their criteria are expected to work to select
heavily absorbed sources.
About 84\% of the objects in their sample are confirmed as 
Compton-thick AGNs and 20\% are newly discovered ones. Thus, the combination of wide field survey data
in the infrared and X-ray bands is promising in the search for heavily obscured AGNs.

In this paper, we combine the infrared all-sky survey data obtained with {\akari} (Murakami et al. 2007)
and the 2XMM
catalog. We construct diagnostic diagrams to classify activity in galaxies and to
search for obscured AGNs. 
We selected 48 candidates for obscured AGNs and analyzed their X-ray spectra.
This paper is organized as follows.
Section 2 describes the selection method of X-ray and infrared sources. 
Diagnostic diagrams to classify the selected sources are presented in Section 3. 
Results of X-ray spectral analysis are shown in section 4.
Section 5 discusses the results and summaries are given in Section 6.
We adopt $H_0$ = 70 km s$^{-1}$, $\Omega_{\rm M}$ = 0.3, and 
$\Omega_\Lambda$ = 0.7 throughout this paper.

\begin{figure}[t]
\begin{center}
\includegraphics[angle=-90,scale=0.35]{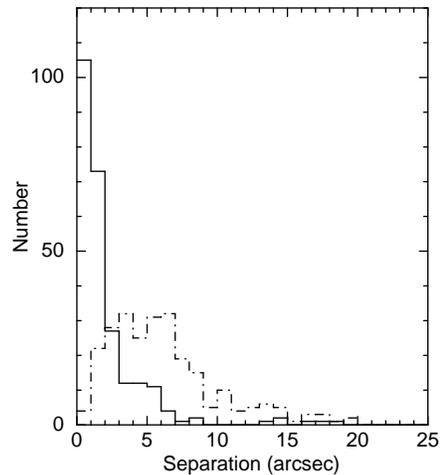}
\caption{
Distribution of the separation between the X-ray and infrared positions. Solid line: 18~$\mu$m sources. Dotted-dashed line: 90~$\mu$m sources.
}
\label{fig:separation}
\end{center}
\end{figure}

\section{THE SAMPLE}


\subsection{XMM-Newton and AKARI catalogs}

We combine two large area survey data in the X-ray and infrared bands.
We used the {\xmm} Serendipitous source catalog Data Release 3 (2XMMi-DR3), which 
contains 262902 unique X-ray sources. The median X-ray flux in 0.2--12 keV is $2.5\times10^{-14}$
\eps \pcm. 
The typical positional uncertainty is 1.\arcsec5
 (1$\sigma$) (Watson et al. 2009).
The entries listed in this catalog (CRs  in 0.2--12 keV and HR4)
are used to create diagnostic diagrams and select candidates for obscured AGNs.
The data from EPIC-PN, which has a larger effective area than those of EPIC-MOS,  was used throughout the analysis.
We use sources located at the Galactic latitude $|b|>10^\circ$ in 2902 observations with usable PN data.
Among these observations, 
2686 and 216 observations are taken with full window and large window modes, respectively.
There are 2062 and 139 unique fields taken with the full window and large window modes, respectively.
The total number of unique EPIC-PN sources at $|b|>10^\circ$  is 150799.
In the following analysis, sources with EPIC-PN counts in 0.2--12 keV greater than 60 counts (60851 unique sources) are used.

{\akari} Point Source Catalogs (PSCs) were used as infrared data. {\akari} surveyed most of the sky
with the two instruments, the Infrared Camera (IRC; Onaka et al. 2007, Ishihara et al. 2010) and the Far-Infrared Surveyor (FIS; Kawada et al. 2007).
The bandpasses of the IRC are centered at 9 and 18~$\mu$m, while the band centers of the FIS
are 65, 90, 140, and 160 $\mu$m. We used 18  and 90 $\mu$m measurements as
mid- and far-infrared data. The 18~$\mu$m is chosen to avoid  silicate features, which may affect continuum measurements,
in the 9~$\mu$m
bandpass of the IRC. The 90 $\mu$m band is used among the fir-infrared bands 
because this band is most sensitive.
We use only data with the quality flag of  FQUAL = 3, which means flux measurements are most
reliable (Yamamura et al. 2010). There are 43865 (18 $\mu$m) and 62326 (90 $\mu$m) sources with FQUAL = 3 located at $|b|>10^\circ$.

\begin{figure}[t]
\begin{center}
\includegraphics[angle=0,scale=0.45]{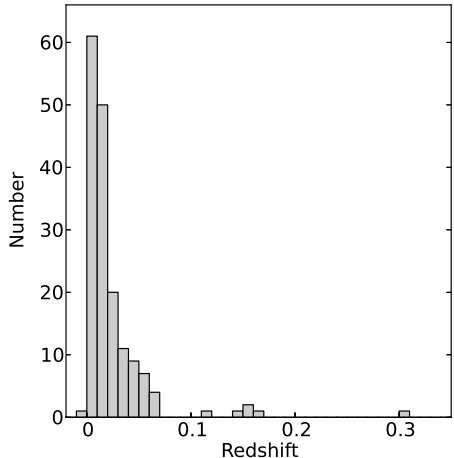}
\caption{
Distribution of redshifts for the matched sample using X-ray, 18, and  90 $\mu$m data shown in Table 1.
}
\label{fig:redshift_all}
\end{center}
\end{figure}

\begin{figure}[t]
\begin{center}
\includegraphics[angle=-90,scale=0.32]{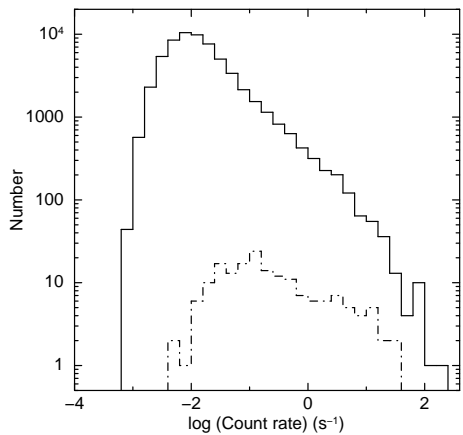}
\caption{
Distribution of X-ray count rates in 0.2--12 keV. Solid line: all the EPIC-PN sources located at $|b|>10^\circ$ with 0.2--12 keV counts greater than 60.
Dotted-dashed line: 173 objects in our sample after matching with the 18 $\mu$m and 90 $\mu$m sources and excluding Galactic and ultraluminous X-ray sources.
}
\label{fig:countrate}
\end{center}
\end{figure}

\subsection{Cross Correlation of the X-ray and Infrared Catalogs}

We compared the positions of X-ray sources in the 2XMM-DR3 and infrared sources in the {\akari} PSCs and 
made a list of X-ray and infrared sources.
We first listed FIS sources within 20\arcsec ~of the X-ray source positions.
If there are multiple sources within the circular region around an X-ray source, the source closest
to the X-ray position is assumed to be a counterpart and is used in the following analysis.
If one source is observed more than twice with {\xmm}, only the data with the largest
number of counts
in 0.2--12 keV is used.
We then matched the IRC and X-ray sources in a similar way using a matching radius of 10\arcsec.
Finally, the FIS and IRC source lists are combined.
The combined list contains 253 sources.
The histograms of the separation between the X-ray and infrared positions are shown in Figure \ref{fig:separation} for 
the
18 and $90~\mu$m sources. The separations for the $18~\mu$m sources are more concentrated 
within a small radius from the X-ray positions compared to the $90~\mu$m sources as expected from
the positional accuracy of the IRC and FIS sources (3$\sigma \approx$6\arcsec and 18\arcsec, respectively, for faint sources).

\begin{figure}[t]
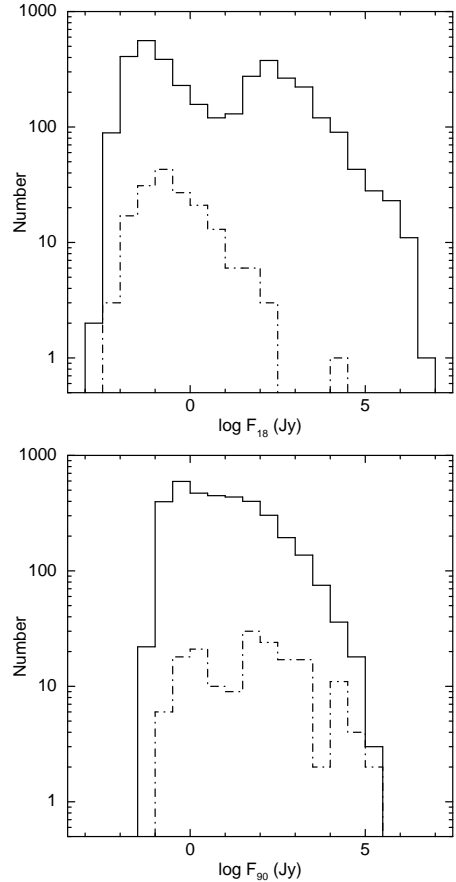

\begin{center}
\includegraphics[angle=-90,scale=0.32]{fig4a.ps}
\includegraphics[angle=-90,scale=0.32]{fig4b.ps}
\caption{
(Upper) Distribution of 18 $\mu$m flux densities. 
(Lower) Distribution of 90 $\mu$m flux densities. 
Solid line: nrearest neighbor 18 $\mu$m sources to 90 $\mu$m sources within 20\arcsec~located at $|b|>10^\circ$.
Dotted-dashed line: 173 objects in our sample after matching with the X-ray sources and excluding the Galactic and ultraluminous X-ray sources.
}
\label{fig:IRflux}
\end{center}
\end{figure}



We searched for the most probable counterpart of the X-ray sources using the 
NASA/IPAC Extragalactic Database (NED) and SIMBAD Astronomical Database.
Stars, \ion{H}{2} regions, planetary nebulae, young stellar objects, and ultraluminous X-ray sources
are excluded from the sample, resulting in 173 objects.
The sample thus obtained is summarized in Table 1, where the {\it XMM} source name, 
alternative name, redshift, infrared flux densities, 
X-ray CRs in 0.2$-$12 keV, and hardness ratios (HR4)
are shown. 
Redshifts are available for 171 of the 173 sources. Their distribution is shown in Figure \ref{fig:redshift_all} as a histogram.
The distribution of X-ray CRs in 0.2--12 keV for all the EPIC-PN sources located at $|b|>10^\circ$ 
with 0.2--12 keV counts greater than 60 and the 171 objects in our sample is shown in Figure \ref{fig:countrate}.
Comparison of these histograms indicates that X-ray brighter objects  tend to have a
possible infrared counterpart detected both in the 18 and 90 $\mu$m bands.
None of X-ray sources fainter than 0.004 counts s$^{-1}$ in 0.2--12 keV are matched with infrared sources. 
We also compared distributions of infrared flux densities. The solid histograms in Figure \ref{fig:IRflux} are
distributions of infrared flux densities for 90 $\mu$m sources having 18 $\mu$m source(s) within 20$\arcsec$
and their nearest neighbor 18 $\mu$m sources (3535 objects in total)
located at $|b|>10^\circ$ with FQUAL = 3. The dotted-dashed histograms are those for the 173 objects 
matched with EPIC-PN sources. A clear difference before and after matching with X-ray sources is shown as
the double peak structure  in the histogram for the 18 $\mu$m sources. The peak at a higher infrared flux density
is likely to be composed of Galactic sources because the peak is more enhanced if we use objects at low Galactic longitude
(e.g., $10^\circ<b<15^\circ$) and because the peak disappeared after excluding Galactic sources in our selection procedure.
The distribution of 90 $\mu$m fluxes for X-ray-matched objects are flatter than that for all the 90 $\mu$m sources with nearby
18 $\mu$m sources. This fact indicates that objects brighter at 90 $\mu$m  tend to have a greater probability of being matched with
an X-ray source.

\begin{figure}[t]
\begin{center}
\includegraphics[angle=0,scale=0.45]{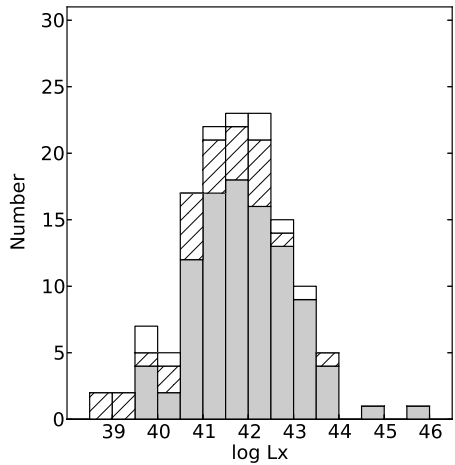}
\caption{Distribution of observed X-ray luminosities in the 2--10 keV band for objects showing AGN activity for
the matched sample using X-ray, 18, and 90 $\mu$m data shown in Table 1  (shaded histogram), 
the 90 $\mu$m sample (hatched histogram), and the 18 $\mu$m (open histogram) sample.
}
\label{fig:Lx}
\end{center}
\end{figure}

\begin{figure}[t]
\begin{center}
\includegraphics[angle=0,scale=0.45]{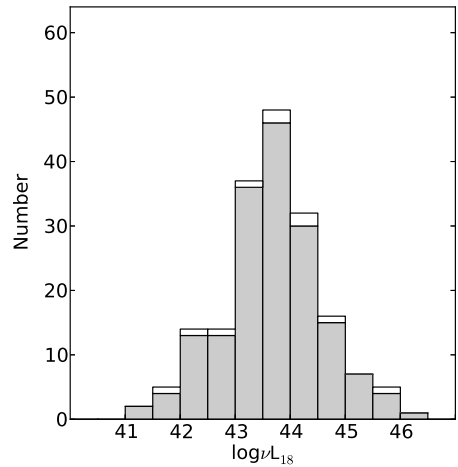}
\hspace{-1.5cm}
\includegraphics[angle=0,scale=0.45]{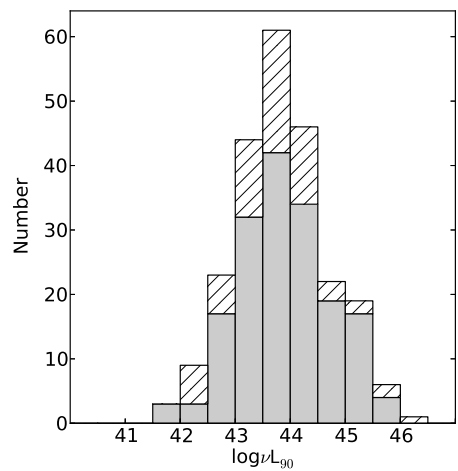}
\caption{Distribution of infrared luminosities $\nu L_\nu$ at 18 $\mu$m (upper) and 90 $\mu$m (lower)
 for the matched sample using X-ray, 18, and 90 $\mu$m data shown in Table 1 (shaded histogram), 
 the 90 $\mu$m sample (hatched histogram),  and the 18 $\mu$m (open histogram) samples.
}
\label{fig:LIR}
\end{center}
\end{figure}

We summarize the X-ray and infrared luminosities for the X-ray-IR matched sample in Table \ref{tb:luminosity}.
The distribution of the luminosities  are shown in Figures \ref{fig:Lx} and \ref{fig:LIR} as shaded histograms.
Distances to nearby objects with a redshift parameter smaller than 0.003 
are taken from the literature shown in Table \ref{tb:luminosity}, except for Mrk 59 for which a redshift-independent distance is not available. 
Distances to objects at $z>0.003$ and Mrk 59 are calculated from source redshifts and the assumed cosmology.
Observed X-ray luminosities in the 2--10 keV band (source rest frame) are obtained from the literature or our analysis presented in section 4 
if X-ray spectra clearly show the presence of an AGN.
Most X-ray fluxes or luminosities are taken from the literature, like for absorption column densities. 
References are given in Table \ref{tb:luminosity} only if a reference different from that for $N_{\rm H}$ is used.
When we use published fluxes in 2--10 keV to calculate luminosities, we made a simple $K$-correction by assuming 
a simple power law with a photon index of 1.8. Since most of the objects in our samples are at a low redshift ($z< \sim 0.3$), 
the assumption on the spectral shape only slightly affects the correction. If a photon index of 1.4 is assumed instead, 
the luminosity would be lower by 10\% at $z=0.3$. This amount is much smaller for lower redshift sources.
In some cases, we obtained observed luminosities from the literature and converted them to our assumed cosmology.
Infrared luminosities ($\nu L_{\nu}$ for 18 and/or 90 $\mu$m) are calculated from the {\it AKARI} measurement of infrared fluxes.
We applied a $K$-correction by assuming a template spectral energy distribution for Seyfert 2s by Poletta et al. (2007). 
Again, because of low redshifts for our sample, the amount of the correction is relatively small.
A correction factor is at most about 10\% for a 18 $\mu$m luminosity at $z=0.3$.

We compiled optical classifications from the literature or spectra in the archives.
The sources are classified into
Seyfert, Low-Ionization Nuclear Emission line Region (LINER), \ion{H}{2} nucleus, transition between LINER and \ion{H}{2}, BL Lac object,
or normal galaxy.
Seyfert, LINER, \ion{H}{2}, and, transition objects are defined based on
the location of the optical emission line ratios ([\ion{N}{2}]$\lambda$6584/H$\alpha$, [\ion{S}{2}]$\lambda\lambda$6716, 6730/H$\alpha$, 
and [{\ion{O}{3}]$\lambda$5007/H$\beta$)
on the excitation diagrams. Among various definitions of the boundary among 
the classes (Baldwin et al. 1981, Veilleux \& Osterbrock 1987, Ho et al. 1997, Kewley et al. 2006)
we adopted criteria of Ho et al. (1997) because many objects in our sample are
contained in the Ho et al. sample in which stellar absorption lines were carefully treated
in measuring emission line fluxes.
Some objects are in the boundary region of two activity classes on the
excitation diagrams, or use of different emission lines results in different classifications.
For such ambiguous cases, both activity classes are shown such as Seyfert/LINER.
The types (1, 1.2, 1.5, 1.8, 1.9, and 2) of Seyferts, LINERs, and
transition objects are also shown if available in the literature. 
The classifications for some objects are not published in the literature, and we
classified their optical spectra from the archives of the Sloan Digital Sky Survey, the 6dF Galaxy survey, or  
the Updated Zwicky Catalog (Falco et al. 1999), if available.
If optical spectra show only absorption lines, they are classified as a normal galaxy.
No classifications are available 
for some of the objects in the sample. They are denoted as ``unclassified.''
These classifications are shown in Table 1 and summarized as histograms in Figure \ref{fig:class}.
In the histograms, Seyfert 1, 1.2, and 1.5
are treated as ``Seyfert 1", while Seyfert 1.8, 1.9, and 2 are regarded as ``Seyfert 2."

\begin{figure}[t]
\begin{center}
\includegraphics[angle=0,scale=0.45]{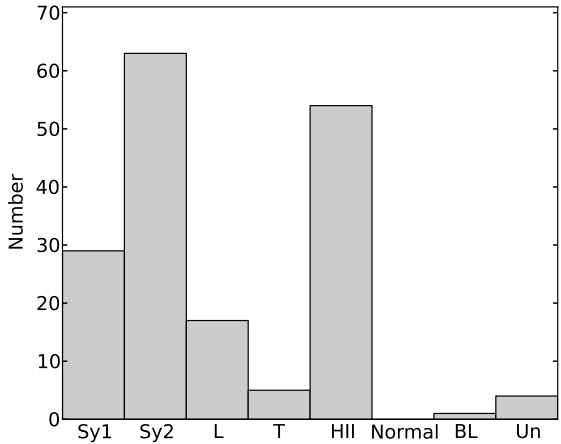}
\caption{
Summary of optical classifications for 
the matched sample using X-ray, 18, and  90 $\mu$m data shown in Table 1.
Sy1: Seyfert 1, 1.2, and 1.5; Sy2: Seyfert 1.8, 1.9, and 2; L: LINER; T: transition object between LINER and HII nucleus; HII: HII nucleus; Normal: normal galaxy; BL: BL Lac object; Un: Unclassified.
}
\label{fig:class}
\end{center}
\end{figure}


Since one of our aims is to search for obscured AGNs, we compiled 
absorption column densities ({\NH}) measured using X-ray spectra from the literature, as
shown in Table 1.
 {\NH} values are shown for Seyferts and LINERs in which X-ray emission is dominated by AGNs.
 Some galactic nuclei classified as \ion{H}{2} show evidence for the presence of an AGN. 
 Their {\NH} values are also shown.
If there are multiple published results, we put priority on the
results of systematic analysis of a large sample, and results based on better
quality of data. For objects showing a signature of heavy absorption exceeding
$\sim10^{24}$ \pcm, we use results based on wide-band spectra covering
hard X-rays above 10 keV whenever possible. 
Objects showing a strong Fe-K emission line with an equivalent width (EW) exceeding 700 eV 
and/or a very flat spectral slope
are regarded as {\NH}$>10^{24}$ {\pcm} even if only X-ray spectra below 10 keV
are available. The boundary of the EW (700 eV) was chosen based on
analysis of X-ray spectra for a large sample of AGNs (Guainazzi et al. 2005a; Fukazawa et al. 2011) and
theoretical predictions (Awaki et al. 1991; Leahy \& Creighton 1993; 
Ghisellini et al. 1994; Ikeda et al. 2009; Murphy \& Yaqoob 2009;
Brightman, \& Nandra 2011a).
If an {\NH} value is not explicitly presented in the literature and if X-ray spectral shape does not 
show a clear signature of absorption, {\NH} is regarded as small as noted in Table \ref{tb:18_90sample}.

\begin{figure}[t]
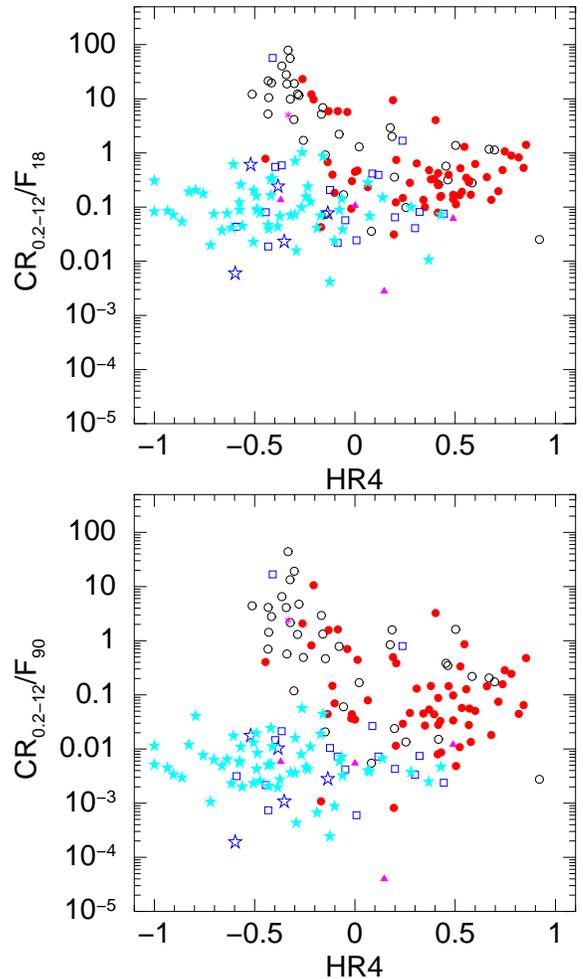

\includegraphics[angle=-90,scale=0.5]{fig8a.ps}
\hspace{5mm}
\includegraphics[angle=-90,scale=0.5]{fig8b.ps}
\caption{
Hardness ratio (HR4) versus count rate (CR$_{0.2-12}$) /  infrared flux ($F_{18}$ or $F_{90}$) ratio diagram. 
Different symbols represent optical classifications; 
open circles: Seyfert 1, 
filled circles: Seyfert 2, 
open squares: LINER, 
open stars: transition object,
filled stars: \ion{H}{2} nucleus, 
asterisk: BL Lac object, 
filled triangles: unclassified. 
(Left) Diagram using
18 $\mu$m flux as infrared flux. (Right) Diagram using 90 $\mu$m flux as infrared flux. 
}
\label{fig:diagram_class}
\end{figure}

\begin{figure}[t]
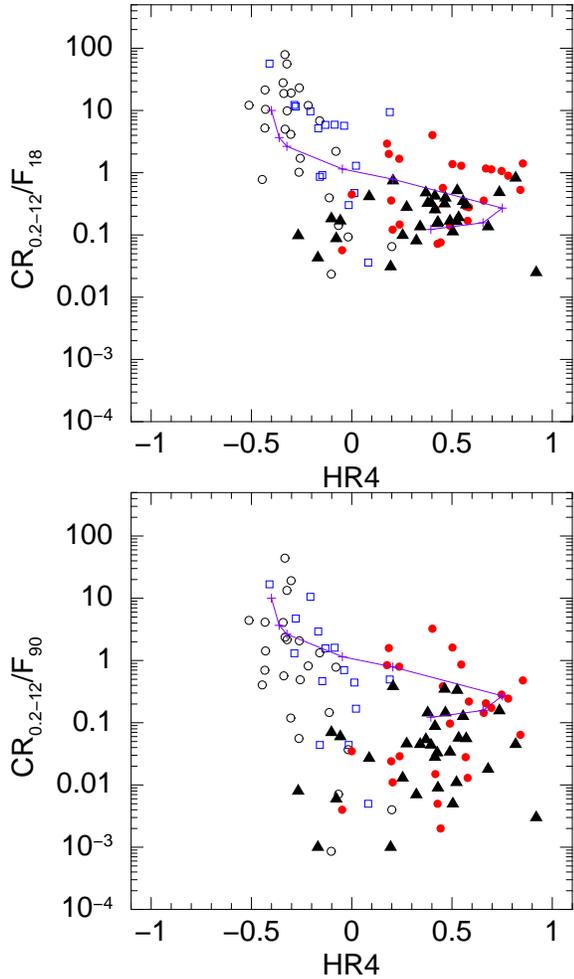

\includegraphics[angle=-90,scale=0.5]{fig9a.ps}
\includegraphics[angle=-90,scale=0.5]{fig9b.ps}
\caption{
Hardness ratio (HR4) versus count rate (CR$_{0.2-12}$) /  infrared flux ($F_{18}$ or $F_{90}$) diagram same as Figure 2 but
only AGNs with measured X-ray spectra are plotted. Different symbols represent \NH ~measured from X-ray spectra; 
open circles: \NH $<10^{22}$ \pcm, 
open squares: $10^{22}$ \pcm $<$ \NH $<$ $10^{23}$ \pcm, 
filled circles:  $10^{23}$ \pcm $<$ \NH $<$ $10^{24}$ \pcm, 
filled triangles: \NH $>10^{24}$ \pcm.
The solid lines are tracks expected for various column densities. EPIC-PN count rate (0.2--12 keV) to infrared flux (18 $\mu$m or 90 $\mu$m)
ratio of 10 is assumed at \NH = 0 cm$^{-2}$. The plus symbols are marked at \NH = (0, 0.5, 1, 5, 10, 50, 100, 150)$\times 10^{22}$ cm$^{-2}$ from the upper-left most to the lowest most point.
See the text for details of the assumed spectral model.
}
\label{fig:diagram_NH}
\end{figure}

\begin{figure}[t]
\includegraphics[angle=-90,scale=0.45]{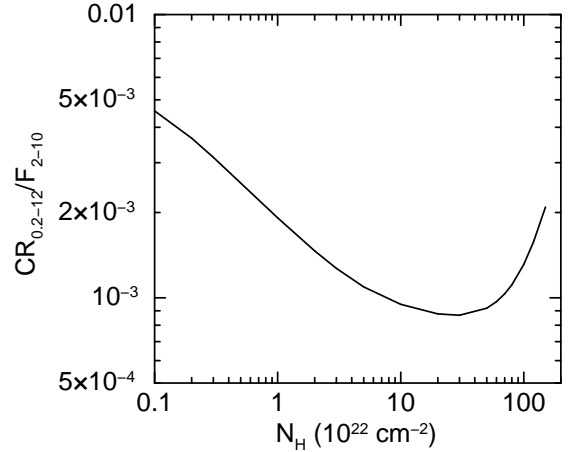}
\caption{
Flux to count rate conversion factor for various \NH~values. 
The expected count rates in 0.2--12 keV for a source with a flux of $10^{-14}$ erg s$^{-1}$ cm$^{-2}$
in 2--12 keV are shown.
See the text for details of the assumed spectral model.}
\label{fig:conversion}
\end{figure}

\begin{figure}[t]
\includegraphics[angle=-90,scale=0.5]{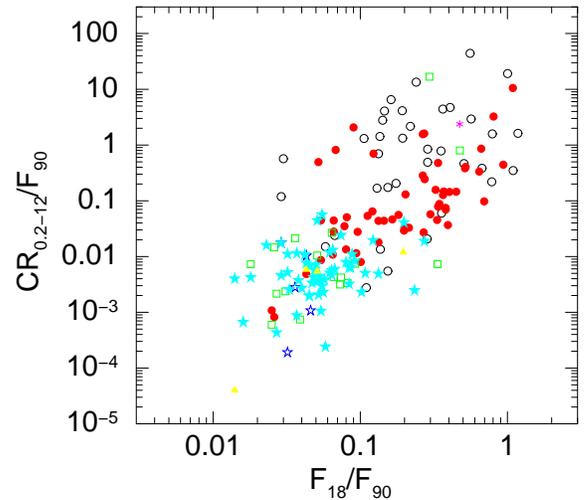}
\caption{
Count rate (CR$_{0.2-12}$) /  infrared flux ($F_{90}$) ratio versus 
infrared flux ratio ($F_{18}/F_{90}$) diagram. The symbols are same as in Figure \ref{fig:diagram_class}.}
\label{fig:IRcolor}
\end{figure}

\section{DIAGNOSTIC DIAGRAMS}

\subsection{Hardness and X-ray/Infrared Ratio}

We first made a diagnostic diagram using the hardness ratio HR4 and the ratio between the X-ray CR in 
0.2--12 keV and the infrared (18 or 90 $\mu$m) flux density.
These diagrams are essentially the same as that used by Severgnini et al. (2012).
We used X-ray CRs instead of X-ray fluxes because X-ray CRs are values directly derived 
from observational data without any assumptions on the X-ray spectral shape.
Our diagrams are shown in the left and right panels of Figure \ref{fig:diagram_class} for 18~$\mu$m and 90 $\mu$m, respectively.
Different symbols are used to represent optical classifications of the activity.
Seyfert 1, 1.2, and 1.5
are denoted as ``Seyfert 1", while Seyfert 1.8, 1.9, and 2 are shown as ``Seyfert 2" in
the diagrams.

Seyfert 1s are located in the upper left part of the diagram, while
Seyfert 2s tend to be located in the lower right.
The larger values of hardness ratios and lower X-ray counts relative to
infrared fluxes of Seyfert 2s are due to the suppression of lower energy X-rays
via photoelectric absorption. 
\ion{H}{2} galaxies are located in the lower left side. 
The location of \ion{H}{2} galaxies indicates that X-rays are
relatively weak compared to AGNs at a given infrared power. 
In order to examine the effect of absorption, 
we defined three groups sorted by \NH~and plotted them using different symbols in Figure \ref{fig:diagram_NH}.
The ordinate and abscissa are same as in Figure \ref{fig:diagram_class}. The open circles, open squares, filled circles, and filled triangles
represent the groups of {\NH} $<10^{22}$ {\pcm}, {\NH} $= 10^{22-23}$ {\pcm}, {\NH} $= 10^{23-24}$ {\pcm}, and {\NH} $>10^{24}$ {\pcm},
respectively. As expected, objects with lower and higher {\NH} tend to be located at around the upper left and 
lower right part of the diagrams, respectively. 


We made tracks of expected hardness ratios and 
X-ray/infrared ratios for various values of {\NH}.
A power-law spectrum with a photon index of 1.8 was assumed as 
incident emission.  One percent  of the incident emission is assumed to appear as 
scattered emission keeping the spectral shape. 
The Galactic absorption of $2\times10^{20}$ \pcm ~is assumed as a representative value.
The expected values were calculated for various intrinsic
column densities from {\NH} = 0 to 1.5$\times10^{24}$ {\pcm}.
The track thus calculated is shown as the solid line in Figure \ref{fig:diagram_NH}, where
the values of CR$_{0.2-12}/F_{18}$ and  CR$_{0.2-12}/F_{90}$ 
are assumed to be 10 when {\NH} = 0 {\pcm}.
The plus signs are marked at {\NH} = $(0, 0.5, 1, 5, 10, 50, 100, 150)\times10^{22}$
{\pcm} from the upper left-most to the lowestr point.
The trend shows that X-ray counts are suppressed 
and that hardness ratio becomes larger for larger {\NH}~values up to \NH $\sim 5\times10^{23}$ {\pcm}.
If {\NH} is larger than \NH $\sim 5\times10^{23}$ {\pcm}, the flux from the 
scattered emission becomes more significant relative to the
absorbed power-law emission, and  the hardness ratio becomes
small.
The above considerations suggest that if objects appeared in the lower right part of the
diagram, they are candidates for obscured AGNs.
Some of the objects in this region  have no published X-ray spectra.
We study the X-ray spectra of such candidates for obscured AGNs
selected from that region 
(CR$_{0.2-12}/F_{90}$ $<0.1$ or
CR$_{0.2-12}/F_{18}$ $<1$)
and HR4 $>-0.1$ in section 4.

A disadvantage of using CRs instead of fluxes is that they depend on the instrument used in observations.
We calculate the conversion factors from X-ray fluxes to observed CRs. The same spectral shape used to 
derive the track shown in Figure \ref{fig:diagram_NH} is assumed. The on-axis response for the EPIC-PN and an integration radius
of 60", which contains 95\% of the total flux, are used. 
The expected CRs for an observed flux in 2--12 keV of $10^{-14}$ erg s$^{-1}$ cm$^{-2}$ are shown as
a function of the absorption column density in Figure \ref{fig:conversion}.

\subsection{Infrared Color and X-ray/Infrared ratio}

Figure \ref{fig:IRcolor} shows a diagram CR$_{0.2-12}/F_{90}$ versus infrared flux ratios ($F_{18}/F_{90}$).
This diagram supplements Figure \ref{fig:diagram_class} in dividing AGNs and \ion{H}{2} nuclei. \ion{H}{2} nuclei
show smaller CR$_{0.2-12}/F_{90}$ and $F_{18}/F_{90}$ ratios compared to AGNs and are located
in the lower left part of the diagram; most \ion{H}{2} nuclei have ratios
CR$_{0.2-12}/F_{90} < 0.05$ and $F_{18}/F_{90} < 0.2$, while only several AGNs 
are found in this region. AGNs tend to show warmer infrared colors compared to \ion{H}{2} nuclei,
and the distributions of the color for Seyfert 1 and Seyfert 2 are almost identical.
These results imply that the dust in AGNs are warmer than those in \ion{H}{2} nuclei
because of the presence of a hard heating source, 
confirming earlier studies (e.g., Wu et al. 2009).
Comparison between type 1 and type 2 AGNs shows that
Mid-IR emission from warm dust near the 
central engine is visible in both types of AGNs. Thus, we confirmed earlier results
based on ground-based observations
(Gandhi et al. 2009), or a combination of \akari~ and hard X-ray surveys
(Ichikawa et al. 2012, Matsuta et al. 2012).

\begin{figure}[t]
\includegraphics[angle=0,scale=0.5]{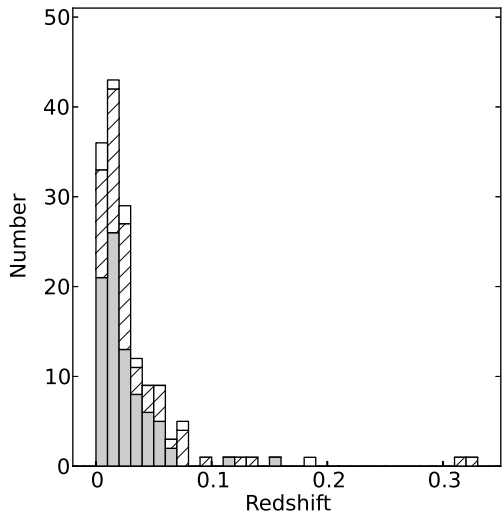}
\caption{
Distribution of redshifts for the 18 + 90 $\mu$m (shaded histogram),
90 $\mu$m (hatched histogram), and 18 $\mu$m (open histogram) samples.\\}
\label{fig:redshift}
\end{figure}

\begin{figure}[t]
\includegraphics[angle=-90,scale=0.33]{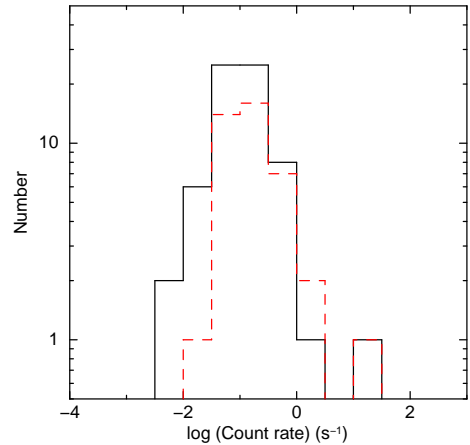}
\caption{Distribution of X-ray count rates in the 0.2--12 keV band for
18 $\mu$m + 90 $\mu$m sample (solid histogram) and Severgnini et al.'s sample (dashed histogram).}
\label{fig:countrate_comparison}
\end{figure}

\begin{figure}[t]
\includegraphics[angle=-90,scale=0.33]{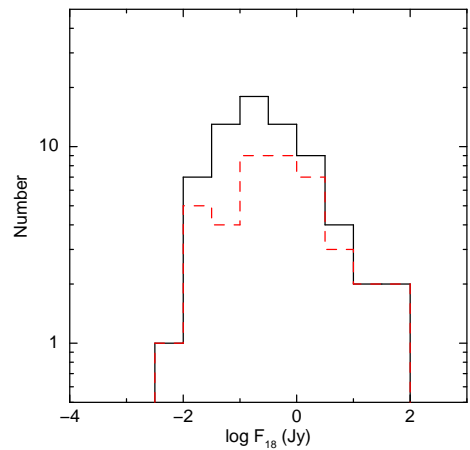}
\caption{Distribution of flux densities at 18 $\mu$m for objects satisfying the condition CR$_{0.2-12}$/$F_{18}<1$
in  18 $\mu$m + 90 $\mu$m sample (solid histogram) and Severgnini et al.'s sample (dashed histogram)}
\label{fig:FIR_comparison}
\end{figure}

\section{X-RAY SPECTRA}

\subsection{The Sample and Data Reduction}

As described in section 3.1, objects in the lower right part of the
diagram HR4 versus CR/infrared flux
ratio are candidates for
obscured AGNs. In order to explore the nature of the candidates, 
we compiled X-ray results from the literature and analyzed X-ray spectra 
of objects for which no published results are available.

We first select objects satisfying (CR$_{0.2-12}/F_{90} < 0.1$  or CR$_{0.2-12}/F_{18} < 1$ ) 
and HR4 $>-0.1$ from the matched sample using X-ray, 18, and 90 $\mu$m.
This sample consists of 85 objects and is denoted as the 18 + 90 $\mu$m sample hereafter.
In addition to this sample, 
we selected objects satisfying
CR$_{0.2-12}/F_{90} < 0.1$ and HR4 $>-0.1$, where
we required the condition FQUAL = 3 only for
90 $\mu$m data to increase the size of the sample. 84 objects are selected by these conditions.
The list of objects in this 90$\mu$m sample is shown in Table \ref{tb:90sample}.
We also selected objects with FQUAL = 3 at 18 $\mu$m, FQUAL $\neq$ 3 at 90 $\mu$m, and
HR4 $>-0.1$. This 18 $\mu$m sample consisting of 10 objects is shown in Table \ref{tb:18sample}.
The probable counterparts, redshifts, optical classifications, and absorption column densities 
determined from X-ray spectra taken from the literature 
are shown in the tables.
Redshifts are available for 83, 62, and 9 objects in the 18 + 90 $\mu$m, 90 $\mu$m, and 18 $\mu$m samples, respectively.
The distributions of the redshifts are shown in Figure \ref{fig:redshift}.
We excluded stellar and off-nuclear sources from the sample as in the
sample shown in Table 1.
We inspected the X-ray data of objects with no published X-ray results
as candidates for detailed studies.
Then the following cases were excluded; 
objects located in bright diffuse emission of a cluster of galaxies or an early-type galaxy, 
in the outskirt of the point spread function of a bright source, on or near the gap between  CCD chips, 
in crowded X-ray source regions such as star-forming regions.

The X-ray and infrared luminosities for the 90  and 18$\mu$m samples are 
summarized in Table \ref{tb:luminosity} and Figures \ref{fig:Lx} and \ref{fig:LIR}, in which both results taken from 
the literature and our own analysis presented in this section are shown.
The $K$-corrections and references for X-ray data and distances are treated 
in the same manner for the sample presented in section 2.2.

We retrieved the data for the sample 
from the {\xmm} Science Archive and examined their spectra.
The {\xmm} Science Analysis Software (SAS) version 13.0.0 and the calibration files
as of 2013 May were used in the data reduction and analysis. 
We first made light curves of a region that does not contain bright sources in 10--12 keV 
to examine background stability and time intervals with high background rates were
excluded. 
We extracted source spectra from a circular region centered at the source position
with a radius of 4$^{\prime\prime}$-- 60$^{\prime\prime}$. 
The extraction radii were determined to achieve good signal to noise ratio and to avoid
nearby sources.
Background spectra were made 
from an off-source region in the same CCD chip and subtracted from the source spectra. 
After data screening and background subtraction, net source counts in 0.2--12 keV for some objects
were found to be lower than 60 counts because of the reduced exposure time. 
Such objects were excluded from the following analysis since their photon statistics
are not sufficient to create spectra of reasonable quality.
The final sample consisting of 48 objects for X-ray spectral analysis and the observation log are summarized 
in Table \ref{tb:Xsample}. The infrared sample from which these objects are taken and Hubble type taken from the HyperLeda database 
(Paturel et al. 2003) are also shown in Table \ref{tb:Xsample}.

The response matrix file and
ancillary  response file were made by using the SAS. The spectra were binned 
so that each bin contains at least one count. 
More channels are binned in the figures shown below for presentation purposes.
A maximum-likelihood method using
the modified version of $C$ statistic (Cash 1979) was employed to fit background-subtracted
spectra. 
Spectral fits were performed with XSPEC version 12.8.0.
The errors represent the 90\% confidence level for one parameter of interest. Errors are not shown 
for the cases that the value of $C$ statistic is much worse than that for the best-fit model.

The Galactic absorption was applied to all the models examined below.
The Galactic absorption column densities (Kalberla et al. 2005)
were obtained by the FTOOL nh and shown in Table \ref{tb:Xsample}.
{\tt phabs} or {\tt zphabs} in XSPEC were used as a photoelectric absorption model.
We examined the presence of an Fe-K emission line at around 6.4 keV by adding a Gaussian component.
The line center energy was left free if the photon statistics were
sufficient to constrain the energy, otherwise 6.4 keV was assumed.
The line width was fixed at a Gaussian $\sigma$ of 10 eV.
For objects with a known redshift, all the model components except for the Galactic absorption
were assumed to be emitted or absorbed at the source redshift. If a redshift is previously unknown 
and an Fe-K emission line is visible in our X-ray spectra, the source redshift was treated as a 
free parameter and was determined from the line, whose central energy was assumed to be 6.4 keV in the source rest frame.
The redshifts  for the X-ray analysis sample 
range from 0.00218 to 0.188 with a median of 0.020, where redshifts determined by an Fe-K line are included.
We assumed a redshift of $z=0$ for all the other cases.
Fe line parameters are not shown if the photon statistics around 6.4 keV are not sufficient and 
if no meaningful constraints on the line is obtained.
The results of the spectral fits described below are summarized in Tables \ref{tb:Xspec} and \ref{tb:Fe}. 
The adopted model is marked with an asterisk in Table \ref{tb:Xspec}.
In these tables, spectral parameters for the best-fit models and models with fewer model components are shown for comparison.
The observed fluxes and
luminosities corrected for absorption (for objects with a known redshift)
in the 2--10 keV band were derived for the best-fit spectral model and models giving a similar quality of fits to the best fit.

\subsection{Results}

We first summarize measurements of absorption column densities taken from the literature, and then
provide detailed results of our own spectral analysis for the 48 objects, for which no X-ray results have been published so far.
The 48 objects analyzed are devided into three groups (1) objects with absorbed spectrum, (2) objects showing very flat continuum and/or
strong Fe-K fluorescent line, and (3) objects showing unabsorbed spectrum, and explained in turn.

\subsubsection{X-ray Results Taken from the Literature}

We compiled the results of X-ray spectral fits for our samples from the literature. 
Since our aim is to find obscured AGNs by combining X-ray and infrared data, 
we tabulated absorption column densities ($N_{\rm H}$) derived from X-ray spectra in Tables \ref{tb:18_90sample}, \ref{tb:90sample}, and \ref{tb:18sample} 
for the 18 + 90 $\mu$m, 90 $\mu$m, and 18 $\mu$m samples, respectively. 
The $N_{\rm H}$ values are classified into three classes ``Unabsorbed'',``Compton-thin'', and ``Compton-thick''
for objects with $N_{\rm H}<10^{22}$ cm$^{-2}$,  
$10^{22}<N_{\rm H}<1.5\times 10^{24}$ cm$^{-2}$,  and 
$N_{\rm H}>1.5\times 10^{24}$ cm$^{-2}$,  respectively. 
If only a lower limit on $N_{\rm H}$ of $1\times10^{24}$ cm$^{-2}$ is presented, such sources are regarded as Compton-thick.
These classifications are shown in the ``X-ray Class'' column in Tables \ref{tb:18_90sample}, \ref{tb:90sample}, and \ref{tb:18sample}. 
The classifications of X-ray absorption are available for 72, 22, and 9 objects
for the 18 + 90 $\mu$m, 90, and 18 $\mu$m samples, respectively.

\subsubsection{Absorbed Spectrum}

X-ray spectra of 16 of the 48 objects we analyzed
show convex shape, which is a signature of 
absorbed emission, implying the presence of an obscured AGN, at energies above $\sim$ 2 keV.
Their spectra are shown in Figure \ref{fig:abs}.
We fitted their spectra by a power-law model absorbed by neutral matter. 
First the photon index was treated as a free parameter, and then a model with a photon index fixed at 1.8
was examined. The result for a free photon index is shown if a meaningful constraint on the index is obtained.

The spectrum of 2MASX~J05430955$-$0829274 is well fitted with this absorbed power-law model accompanied by an
emission line at 6.4 keV.
The rest of the objects show additional emission at energies below a few keV.
We tried to model this emission by power law, APEC thermal plasma (Smith et al. 2001), or a combination of both.
For the models with one power-law component, the photon index was left free or fixed at 1.8.
The photon indices of the heavily absorbed power law, which dominates hard emission, and 
the additional power law representing the soft part of the spectra were fixed at 1.8 for models containing 
two power-law components.
A common photoelectric absorption model, which represent the absorption in the host galaxies,
was applied to both of the power-law components. 
The APEC component was assumed to be absorbed only by the Galactic column.
The spectra of the two objects 2MASX~J05052442$-$6734358 and NGC 5689 are represented by the two power-law model.
Other objects show excess emission around 0.6--1.0 keV, which implies the presence of Fe-L emission lines from hot plasma, 
and the APEC model was used to express this feature. 
An APEC + absorbed power-law model describes the spectra of  
UGC 959 and IC 5264. 
The spectra of the rest of the objects were fitted with a combination of two power-law and APEC components. 
IRAS 03156$-$1307 required two temperature APEC components in addition to the two power-law components.

The best-fit column densities for the heavily absorbed power-law component are 
\NH $\approx 2\times10^{22} - 1.4\times10^{24}$ \pcm, 
which is a range expected for obscured AGNs. 
The column densities of the two objects IRAS~01356$-$1307 and NGC 2611 exceed $1\times10^{24}$ {\pcm} 
and the effect of Compton scattering cannot be neglected. 
Therefore, we multiplied the {\tt cabs} model in XSPEC ($e^{-\sigma_{\rm T} N_{\rm H} }$), where $\sigma_{\rm T}$ is 
the Thomson scattering cross section. 
Although the energy dependence of the cross section is not taken into account,
this model approximates the shape of the  continuum transmitted through Compton-thick matter  (Ikeda et al. 2009). 
This model affects only the normalization of the heavily absorbed power-law component. 

The results of the spectral fits are summarized in Table \ref{tb:Xspec}, 
and the adopted models and data/model ratios are shown in Figure \ref{fig:abs}. 
If two or more models provide similar $C$ statistics, we adopt models with the best-fit photon index in the range of 1.5--2.1, 
which is typical for X-ray spectra of AGNs.
The results of spectral fits to the Fe-K line are shown in Table \ref{tb:Fe}. 
The improvement of the $C$ statistic by adding a Gaussian line is also shown in Table \ref{tb:Fe}.
An Fe-K emission line is seen in the spectra of 2MASX~J05052442$-$6734358, 2MASX~J05430955 $-$0829274, and ESO~205$-$IG003. 
A hint of Fe-K emission is seen in SDSS J085312.35+162619.4.
The photon statistics are poor around 6.4 keV in IRAS 03136$-$1307 and Fe line parameters are not shown
for this object.
The improvement of the $C$ statistic for other objects is small for one additional parameter (normalization of a Gaussian).

\subsubsection{Flat Continuum and Strong Fe-K Emission Line}

Ten objects show a flat continuum 
at energies above a few keV and/or a strong Fe-K emission line at around 6.4 keV.
The spectral shape is much flatter than that typically observed in AGNs and
implies that the spectrum is a combination of heavily absorbed and less absorbed power laws with
typical photon indices for AGNs (two power-law model) or that the spectrum is reflection-dominated. 
Therefore, we examined continuum models for these two cases. 
We assumed a  common photon index of 1.8 for the two power-law model.
The {\tt pexrav} model in XSPEC was used to represent
a continuum reflected from cold matter (Magdziarz \& Zdziarski 1995). The incident spectrum is assumed to be a power law 
with a photon index of 1.8 and an exponential cutoff at 300 keV. The reflection scaling factor ({\tt rel\_refl})
was set to $-1$ to represent reflected emission alone. The inclination angle of the reflector was assumed to be 60$^\circ$, where
the inclination of $0^\circ$ corresponds to face on.

The spectrum of 2MASX~J05391963$-$0726190 is fitted either by a pure reflection model or a two power-law model.
A pure reflection model does not describe the spectra of  IRAS 00517+4556, 
2XMM~J052555.5$-$661038, and 
2XMM~J184540.6$-$630522.
The continuum of 2XMMi~J184540.6 $-$630522 is fitted by a flat power-law model or a combination of reflection and slightly absorbed 
power law.  A combination of reflection and virtually unabsorbed power law or a two power-law model 
represent the spectra of IRAS 00517+4556 and 2XMMi~J052555.5$-$661038.

The six objects, NGC 1402, IC 614, 2MASX~J11594382$-$2006579, NGC 6926, 2MASX~J23404437 $-$1151178, and 
NGC 7738, show excess emission around 0.6--1.0 keV suggesting the presence of emission from optically thin plasma.
We used an APEC plasma model to represent this feature.
An APEC + reflection model describes the continua of IC 614, 2MASX J11594382$-$2006579, and NGC 6926.
An APEC + absorbed power-law model also provided a similar quality of fit to 2MASX J11594382$-$2006579.
2MASX~J23404437$-$1151178 and NGC 7738 require an additional power-law component, where we assumed a common 
{\NH} for the reflection and power-law components since the quality of the data is not sufficient to constrain {\NH} values for 
these components separately.
The spectrum of NGC 1402 is not very flat, and a combination of APEC 
and lightly absorbed power law represents the shape of the continuum.

The six  objects, NGC 1402, IC 614, 2XMMi~J184540.6$-$630522, NGC 6926, 2MASX~J23404437 $-$1151178, and NGC 7738,
show an emission line at around 6.4 keV in the source rest frame with an EW in the range of 1.1--4.6 keV. 
An Fe-K line at 6.4 keV is not detected in other objects except for a weak hint of a line in NGC 4713. 
An emission-line-like excess at around 7.0 keV is seen
in the spectrum of IRAS 12596$-$1529. If this line is assumed to be from H-like Fe at 6.97 keV, 
the $C$ statistic is improved by $\Delta C = 5.5$ for one
additional parameter (normalization of a Gaussian). 
2MASX~J11594382$-$2006579 shows a line-like emission at $\approx 5.8$ keV. 
If this emission is an Fe-K line at 6.4 keV in the source rest frame, the redshift is estimated 
to be $0.112^{+0.019}_{-0.018}$, though the improvement of the fit is only $\Delta C = 3.0$
for two additional parameters (normalization of a Gaussian and source redshift).
Limits on the EW of Fe-K line at 6.4 keV were derived for objects with sufficient counts
around 6.4 keV.

The results of the spectral fits are summarized in Tables \ref{tb:Xspec} and \ref{tb:Fe}. 
The observed spectra, adopted models, 
and data/model ratios are shown in Figure \ref{fig:flat}.
If two or more models give fits of similar quality, we adopt one or two models 
as the most appropriate ones satisfying the following conditions:
(1) the best-fit photon index is in the range of 1.5--2.1 and 
(2) the constraint on the EW on an Fe-K fluorescent line at 6.4 keV is consistent with the
best-fit $N_{\rm H}$ as observed in obscured AGNs (Guainazzi et al. 2005a, Fukazawa et al. 2011).

\begin{figure*}[t]
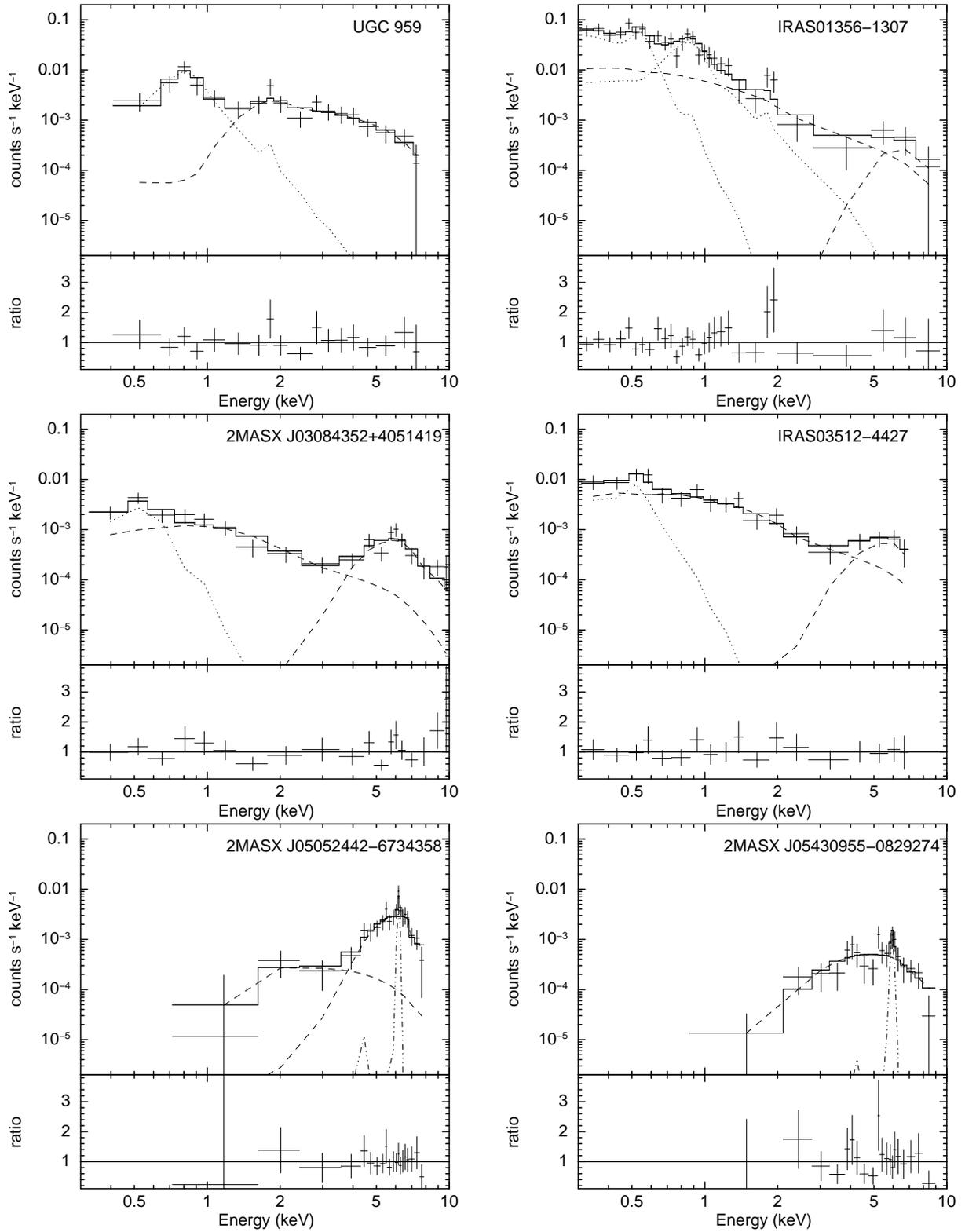

\centering
\includegraphics[angle=-90,scale=0.39]{fig15a.ps}
\includegraphics[angle=-90,clip,scale=0.39]{fig15b.ps}
\includegraphics[angle=-90,scale=0.39]{fig15c.ps}
\includegraphics[angle=-90,scale=0.39]{fig15d.ps}
\includegraphics[angle=-90,scale=0.39]{fig15f.ps}
\includegraphics[angle=-90,scale=0.39]{fig15g.ps}
\caption{EPIC-PN spectra of objects showing absorbed continuum. 
(Upper panel) Data (crosses) and adopted best-fit model (solid histogram). 
Spectral components are shown as dashed, dotted, dot-dot-dot-dashed lines.
(Lower panel) Data/Model ratio. }
\label{fig:abs}
\end{figure*}
\addtocounter{figure}{-1}
\begin{figure*}[h]
\centering
\includegraphics[angle=-90,scale=0.39]{fig15h.ps}
\includegraphics[angle=-90,scale=0.39]{fig15i.ps}
\includegraphics[angle=-90,scale=0.39]{fig15j.ps}
\includegraphics[angle=-90,scale=0.39]{fig15k.ps}
\includegraphics[angle=-90,scale=0.39]{fig15l.ps}
\includegraphics[angle=-90,scale=0.39]{fig15m.ps}
\caption{Continued.}
\end{figure*}
\addtocounter{figure}{-1}
\begin{figure*}[h]
\centering
\includegraphics[angle=-90,scale=0.39]{fig15n.ps}
\includegraphics[angle=-90,scale=0.39]{fig15o.ps}
\includegraphics[angle=-90,scale=0.39]{fig15p.ps}
\includegraphics[angle=-90,scale=0.39]{fig15q.ps}
\caption{Continued.}
\end{figure*}

\begin{figure*}[h]
\centering
\includegraphics[angle=-90,scale=0.39]{fig16a.ps}
\includegraphics[angle=-90,scale=0.39]{fig16b.ps}
\includegraphics[angle=-90,scale=0.39]{fig16c.ps}
\includegraphics[angle=-90,scale=0.39]{fig16d.ps}
\includegraphics[angle=-90,scale=0.39]{fig16e.ps}
\includegraphics[angle=-90,scale=0.39]{fig16f.ps}
\caption{EPIC-PN spectra of objects showing flat continuum and/or strong Fe-K emission.
}
\label{fig:flat}
\end{figure*}
\addtocounter{figure}{-1}
\begin{figure*}[h]
\centering
\includegraphics[angle=-90,scale=0.39]{fig16g.ps}
\includegraphics[angle=-90,scale=0.39]{fig16h.ps}
\includegraphics[angle=-90,scale=0.39]{fig16i.ps}
\includegraphics[angle=-90,scale=0.39]{fig16j.ps}
\caption{Continued.}
\end{figure*}

\begin{figure*}[t]
\centering
\includegraphics[angle=-90,scale=0.39]{fig17a.ps}
\includegraphics[angle=-90,scale=0.39]{fig17b.ps}
\includegraphics[angle=-90,scale=0.39]{fig17c.ps}
\includegraphics[angle=-90,scale=0.39]{fig17d.ps}
\includegraphics[angle=-90,scale=0.39]{fig17e.ps}
\includegraphics[angle=-90,scale=0.39]{fig17f.ps}
\caption{EPIC-PN Spectra of unabsorbed objects.}
\label{fig:ambiguous}
\end{figure*}
\addtocounter{figure}{-1}
\begin{figure*}[t]
\centering
\includegraphics[angle=-90,scale=0.39]{fig17g.ps}
\includegraphics[angle=-90,scale=0.39]{fig17h.ps}
\includegraphics[angle=-90,scale=0.39]{fig17i.ps}
\includegraphics[angle=-90,scale=0.39]{fig17j.ps}
\includegraphics[angle=-90,scale=0.39]{fig17k.ps}
\includegraphics[angle=-90,scale=0.39]{fig17l.ps}
\caption{Continued}
\end{figure*}
\addtocounter{figure}{-1}
\begin{figure*}[t]
\centering
\includegraphics[angle=-90,scale=0.39]{fig17m.ps}
\includegraphics[angle=-90,scale=0.39]{fig17n.ps}
\includegraphics[angle=-90,scale=0.39]{fig17o.ps}
\includegraphics[angle=-90,scale=0.39]{fig17p.ps}
\includegraphics[angle=-90,scale=0.39]{fig17q.ps}
\includegraphics[angle=-90,scale=0.39]{fig17r.ps}
\caption{Continued}
\end{figure*}
\addtocounter{figure}{-1}
\begin{figure*}[h]
\centering
\includegraphics[angle=-90,scale=0.39]{fig17s.ps}
\includegraphics[angle=-90,scale=0.39]{fig17t.ps}
\includegraphics[angle=-90,scale=0.39]{fig17u.ps}
\includegraphics[angle=-90,scale=0.39]{fig17v.ps}
\caption{Continued}
\end{figure*}

\clearpage

\subsubsection{Unabsorbed Spectrum}


The rest of  the objects  do not show a signature of heavy absorption. 
We applied an absorbed power-law model to the spectra.
The photon index $\Gamma$ was first treated as a free parameter. If the photon index was not well constrained, 
$\Gamma = 1.8$ was assumed.
Good fits were obtained 
for 12 objects (2XMM J004330.4$-$180107, 2MASX~J02253645$-$0500123, 
A426[BM99]183, AKARI~J0531228+120057, 
2XMMi J053512.2$-$690009, CXO J054532.6$-$001129, 
MCG +01$-$27$-$029, UGC 6046, NGC 3953, 
NGC 5132, 2XMM~J222942.7 $-$204607, and  NGC 7617).
An APEC plasma model was also examined instead of power law, and 
similar quality of fits were obtained for 
A426[BM99]183, AKARI~J0531228+120057, 
2XMMi J053512.2$-$690009, 
UGC 6046, NGC 3953, and 2XMM~J222942.7$-$204607.
The spectra of UGC 587 and CGCG~009$-$061A appear very soft, and an APEC model 
with a temperature of 0.3--1 keV provided a good description of the data.

The single-component models do not fit the spectra of 
eight objects (NGC 35, ESO~264$-$G032, 
NGC 4559A, NGC 4696B, 
NGC 4713, IRAS 12596$-$1529,
NGC 5350, and 2MASX J14341353 +0209088). Although the spectra of 
NGC 5132 and NGC 7617 are fitted by a power-law model, the resulting photon indices
are very steep (3.3 and 2.8, respectively) and may indicate the presence of a soft component.
We therefore examined a two component-model consisting of APEC plasma and absorbed power law
with a fixed photon index of 1.8.
This model describes the spectra of all but NGC 4696B. The addition of a second APEC component
well fits the spectrum of NGC 4696B. 

Two objects (NGC 4713 and IRAS 12596$-$1529) show a hint of an emission line at 6--7 keV.
The improvements of the $C$ statistic are 5.3 and 5.5 for one additional parameter (normalization of a Gaussian)
if the line center energies of 6.4 keV and 6.97 keV were assumed for NGC 4713 and IRAS 12596$-$1529,
respectively.
Other objects do not show an Fe-K emission line feature. The upper limits on the EW of a Gaussian line at 6.4 keV
were derived for objects with sufficient photon statistics around an Fe-K line and summarized in Table \ref{tb:Fe}.
The results of the fits are summarized in Table \ref{tb:Xspec}.
The observed spectra and the adopted models are shown in Figure \ref{fig:ambiguous}.

\section{Discussion}

\subsection{Selection of Obscured AGNs}

We made diagnostic diagrams using X-ray CRs, X-ray hardness, and infrared fluxes, which can be used to
select candidates for obscured  AGNs. By accumulating the published results of X-ray spectral analysis,
we found that the regions satisfying
(CR$_{\rm 0.2-12}/F_{18}<1.0$ or
CR$_{\rm 0.2-12}/F_{90}<0.1$)
 and HR4$>-0.1$ are the loci for obscured AGNs.
We analyzed the X-ray spectra of 48 objects with X-ray counts greater than 60, for
which no X-ray spectra are published.
Of these, 26 show a signature of absorbed X-ray spectra.
Their classifications of X-ray absorption as Compton-thin or Compton-thick are summarized in Table \ref{tb:Xspec}.
The 16 objects analyzed in section 4.2.2 show a spectral curvature indicative of 
the continuum absorbed by  \NH$\approx10^{22-24}$ \pcm. Their best-fit \NH~values are
in the range of $3\times10^{22} - 1.4\times10^{24}$ \pcm, which are typically observed in Seyfert 2 galaxies.
The two objects showing the largest best-fit \NH~
($1.05\times10^{24}$ cm$^{-2}$ for IRAS 01356$-$1307 and $1.4\times10^{24}$ cm$^{-2}$ for NGC 2611)
are regarded as Compton-thin in the discussion below 
since these \NH ~values are slightly below the boundary of Compton-thin/thick
column densities ($1.5\times10^{24}$ cm$^{-2}$; Comastri 2004).
Note, however, 
 that the boundary value of \NH~between Compton-thin/thick depends on the assumed abundance of the absorber
(Yaqoob et al. 2010).
The 10 objects analyzed in section 4.2.3 show a flat continuum and/or a strong Fe-K emission line. 
Six of these objects are most likely to be Compton-thick AGNs judging from their flat continuum and
strong Fe-K emission line. 
The Fe-K emission line is not significant in four 
(IRAS 00517+4556, 2MASX~J05255807$-$6610523, 
2MASX~J05391963$-$0726190, and
2MASX~J11594382$-$2006579) of the 10 objects, and
their flat continuum could be interpreted as either a reflection-dominated spectrum or
a combination of mildly absorbed ($4\times10^{22}-4\times10^{23}$ \pcm) and less absorbed components.
These objects are tentatively regarded as Compton-thick in the following discussion and denoted as ``Thick?" in
the X-ray class column of Table \ref{tb:Xspec}.
Thus, 26 objects in total are most likely obscured AGNs, for which X-ray signatures of the presence of AGNs
are reported for the first time in this paper. 
The 22 objects analyzed in section 4.2.4, on the other hand, show no clear evidence for the presence of obscured AGNs.

By combining the X-ray results taken from the  literature and our own analysis, 
Table \ref{tb:numbers} summarizes the number of objects in the 18 + 90 $\mu$m, 90, and 18 $\mu$m samples,
objects with X-ray measurements of absorption column densities, and unabosrbed/Compton-thin/Compton-thick objects.
Usable X-ray data are available for 84\% of the combined sample.
Among the 151 objects with X-ray measurements, 113 (75\%) are absorbed by Compton-thin or Compton-thick matter. 
If only sources securely detected at 18 $\mu$m are used, 82 of 89 objects (92\%) are absorbed.

The MIR band is better for detecting emission from warm dust heated by an AGN compared to 
the far-infrared band. Therefore,  
we first discuss the success rate to find obscured AGNs selected by the criteria using 18 $\mu$m data.
Among our {\it XMM}+{\it AKARI} sample, 79 objects satisfy the conditions
CR$_{\rm 0.2-12}/F_{18}<1.0$ and HR4 $>-0.1$.
The absorption column densities of 44 and 29 objects among them are
in \NH $= 1\times10^{22} - 1.5\times10^{24}$ \pcm~ and larger than $1.5\times 10^{24}$ \pcm, respectively.
The detection rate of AGNs absorbed by \NH$>10^{22}$ \pcm ~is 73/79 $\approx$92\%, which demonstrates
the efficiency of our criteria finding obscured AGNs.


Severgnini et al. (2012) applied a similar selection technique to the sample derived from the {\it IRAS} 
point source and 2XMM catalogs. They used the
conditions of $F(2-12~{\rm keV})/\nu_{25} F_{25} < 0.02$ and HR4 $>-0.2$ and selected
43 candidates for Compton-thick AGNs. 
Of the 43 AGNs, 40 are in common with our sample of obscured AGNs. 
Two objects (IRAS 04507+0358 and 3C321) are not selected in our sample since EPIC-PN data are not 
available for them. One object (NGC 5194) has an HR4 value of $-0.11$ that is slightly smaller than our 
adopted boundary (HR4 $>-0.1$) but satisfies Severgnini et al.'s criterion (HR4 $>-0.2$).

In Severgnini et al.'s sample, 32 of 43 are confirmed to be Compton-thick AGNs.
The classification of Compton thickness for four objects are model-dependent, and seven are Compton-thin.
If the conversion between {\it IRAS} 25 $\mu$m and {\it AKARI} 18 $\mu$m flux densities of equation (3) in Ichikawa et al. (2012) is assumed,
their  criterion translates into CR$_{0.2-12}/F_{18}<$ 0.24 and 0.64 for
assumed \NH ~of $5\times10^{23}$ {\pcm} and $1\times10^{22}$ \pcm, respectively,
where the same spectral shape  as in section 3.1 was used.
Thus, our criteria (CR$_{\rm 0.2-12}/F_{18}<1.0$) cover X-ray brighter objects 
relative to MIR fluxes, and this might result in the smaller Compton-thick fraction (44/79) derived for our sample.

Another difference between Severgnini et al.'s and our selection criteria is the limited X-ray and infrared flux levels.
Severgnini et al. selected objects with an X-ray flux in 4.5--12 keV larger than $1\times10^{-13}$ erg s$^{-1}$ cm$^{-2}$.
This flux corresponds to $7\times10^{-14}$ erg s$^{-1}$ cm$^{-2}$ in 2--10 keV, 
if \NH = $5\times10^{23}$ {\pcm} and the spectral shape used in section 3.1 are assumed.
The fluxes in 2--10 keV of all but one in our sample are larger than this flux. The flux of one object 
(2MASX~J05391963$-$0726190) is $\sim4\times10^{-14}$ erg s$^{-1}$ cm$^{-2}$ in 2--10 keV, which 
is only slightly below the flux limit. 
We also compared  the distributions of X-ray count rates in 0.2--12 keV. 
The solid histogram in Figure \ref{fig:countrate_comparison}  is the distribution for 68 obscured AGN candidates 
satisfying the 18  or 90 $\mu$m criteria selected from Table 1.
The distribution of the CRs for the Compton-thick candidates in Severgnini et al. (2012) 
is shown as a dashed histogram in the same figure. Our sample contains more X-ray fainter objects
compared to Severgnini et al.'s sample. Therefore, the X-ray flux limit likely explains 
the difference in the Compton-thick fraction in part.
The distributions of 18 $\mu$m fluxes for the 70 objects 
satisfying the condition CR$_{\rm 0.2-12}/F_{18}<1$ in the 18  + 90 $\mu$m sample
and Severgnini et al's Compton-thick
candidates are also compared in Figure \ref{fig:FIR_comparison}.
We compiled 18 $\mu$m fluxes measured with  {\it AKARI} IRC for the latter.
Six sources do not have 18$\mu$m data, and their {\it IRAS} 25 $\mu$m fluxes are converted to 18 $\mu$m fluxes 
by using the equation (3) in Ichikawa et al. (2012). Although the lower bounds of the distribution are similar,
our sample contains a somewhat larger number of objects at fainter infrared fluxes. 
In summary, the combination of the larger numbers of X-ray faint or infrared faint objects compared to Severgnini et al.'s
is likely to be a reason why we were able to find new obscured AGNs not included in Severgnini et al.'s sample.

We also used criteria using far infrared fluxes at 90 $\mu$m, CR$_{\rm 0.2-12}/F_{90}<0.1$ and HR4 $>-0.1$. 
The contribution from cold dust heated by stellar processes in the host galaxy is likely to be significant in
the far infrared band unless  an AGN overwhelms the emission from the host galaxy. 
Therefore, a low CR$_{\rm 0.2-12}/F_{90}$
ratio does not necessarily mean that X-ray emission is weak relative to the infrared
because of the heavy absorption of an AGN. 
We thus expect the criteria using far infrared
are less efficient to select obscured AGNs compared to the MIR selection.
On the other hand, the combination of X-ray hardness and a low X-ray/IR ratio
provides us with a chance to find AGNs buried in starburst activity.
There are 83 objects satisfying the conditions CR$_{\rm 0.2-12}/F_{90}<0.1$,
HR4 $>-0.1$, and not securely detected in the 18 $\mu$m band.
The presence of AGNs absorbed by a column density greater than $10^{22}$ cm$^{-2}$ is reported for
31 objects in the literature or in this paper. 
The \NH~values for 21 and 10 objects are in the range of $1\times10^{22}-1.5\times10^{24}$  \pcm~and 
larger than $1.5\times10^{24}$ \pcm, respectively, where three objects classified as ``Thick?" are regarded as Compton-thick.
There is  no indication of absorption in excess of $10^{22}$ cm$^{-2}$ in 31 objects.
No \NH~measurements are available for the rest of the objects (19 objects).
Therefore, the detection rate of obscured AGNs using the far infrared criteria is 31/62=50\% for the sample with \NH~ measurements.

The 22 objects analyzed in section 4.2.4 show no clear evidence for the presence of an obscured AGN.
This result apparently contradicts our selection using hard X-ray spectra measured by the hardness ratio of
HR4 $>-0.1$. A possible reason for this contradiction is the faintness of the sources in the hard X-ray band.
The net counts in the 4.5--12 keV for 15 among the 22 objects are less than 42 according to the 2XMMi
catalog. The uncertainties of the hardness ratios are very large for such faint objects and sources with
an unobscured spectrum could be chosen by our criteria. In one case (UGC 587), almost no photons 
are seen above 2 keV in our X-ray spectrum, but the 4.5--12 keV count in the 2XMMi catalog is 110 counts.
This case is likely to be due to a combination of very low real X-ray counts and uncertainties in background
estimation. The 4.5--12 keV counts for rest of the  objects are in the range of 
120--250 and their HR4 values are around 0.0, which is relatively soft among the objects selected by
our criteria. The inspection of their spectra indicates that the HR4 values are
reliable and consistent with the observed unabsorbed spectra within the errors.

\subsection{Obscured AGNs Outside the Selection Criteria}

While our selection criteria efficiently select obscured AGNs as discussed in the previous subsection, 
there are some obscured AGNs located outside our criteria in Figure \ref{fig:diagram_NH}.
We examine the nature of these outliers using the 18 + 90 $\mu$m sample.
None of 29 objects absorbed by $N_{\rm H}>10^{24}$ cm$^{-2}$ are located 
in the region CR$_{\rm 0.2-12 keV}$ /$F_{18}<1$.
Of the 26 objects in the range of $N_{\rm H} = 10^{23-24}$ cm$^{-2}$, 9 have 
CR$_{\rm 0.2-12 keV}$ /$F_{18}>1$. This ratio scatters from object to object, 
and the number of objects with CR$_{\rm 0.2-12 keV}$/$F_{18}>1$ depends on
the choice of the boundary. We set the criteria to efficiently select 
more absorbed objects ($>10^{24}$ cm$^{-2}$), which resulted in missing
some moderately absorbed objects ($N_{\rm H} = 10^{23-24}$ cm$^{-2}$) in 
our criteria.

Seven out of 29 sources absorbed by $N_{\rm H}>10^{24}$ cm$^{-2}$
and 12 out of 26 absorbed by $N_{\rm H}=10^{23-24}$ cm$^{-2}$ are outside our criteria
using 90 $\mu$m data. 
Inspection of Figure \ref{fig:diagram_NH} (left and right) clearly shows that scatters 
in X-ray to infrared ratios are much larger if 90 $\mu$m data are used. 
A possible reason is that there is a wide range of the contribution of 
infrared emission from relatively cold dust
heated by sources other than AGNs depending on the nature of host galaxies.
While the scatters naturally worsen the success rate to find obscured AGNs, 
there is a higher probability of finding AGN activity hidden behind starbursts.

Among 55 sources with $N_{\rm H}>10^{23}$ cm$^{-2}$, 
three objects do not satisfy the condition HR4 $>-0.1$.
In these objects (NGC 3690, HR4=$-0.267$; 
Mrk 1, HR4=$-0.101$; NGC 2623, HR4=$-0.169$),
there is a considerable contribution from soft thermal emission to the band pass
used to calculate HR4, where the 2.0--4.5 keV band is used as the soft band.
Since their absorption column densities are above $10^{24}$ cm$^{-2}$,
their X-ray spectra of AGN component below 10 keV are reflection-dominated and the low-energy
cutoff due to photoelectric absorption is not clearly seen. The combination of 
the significant contribution of soft thermal emission and the absence of 
absorbed direct emission results in the relatively small HR4 values.
If starburst activity coexists with obscured AGNs, soft thermal emission from starburst
contributes to X-rays. In the infrared band, emission from cool dust associated with
star formation activity results in smaller X-ray to 90 $\mu$m ratios. Such cases
tend to be missed if our criteria are applied.

\subsection{Optically Elusive AGNs}


There are several objects classified as an \ion{H}{2} nucleus showing a relatively large hardness ratio.
In the 18 + 90 $\mu$m sample, the hardness ratios HR4 of nine objects 
(IRAS 01173+1405, NGC 695, NGC 3877, IRAS 12550$-$2929, IRAS 12596$-$1529, NGC 5253, IRAS 20551$-$4250, 
IRAS 23128$-$5919, and NGC 7738)  are greater than $-0.1$. The presence of an AGN is known in
NGC 695, IRAS 12550$-$2929, IRAS 20551$-$4250, and IRAS 23128$-$5919
(Brightman \& Nandra 2011a, Severgnini et al. 2012) and the latter three show
significant absorption. A signature of the presence of a Compton-thick AGN is 
clearly seen in our spectrum of  NGC 7738. All of these AGNs satisfy the 
conditions CR$_{0.2-12}/F_{18} < 1$ and CR$_{0.2-12}/F_{90} < 0.1$.

Fifteen objects classified as  \ion{H}{2} nuclei in the 90 $\mu$m sample
have HR4 $>-0.1$. Obscured AGNs are found in NGC 1402, 
SDSS  J085312.35+162619.4, IRAS 13443+0802NE, and IC 5264.
The former object is Compton-thick and the latter three are Compton-thin.
Seven objects show no clear indication of the presence of an obscured AGN
(section 4.2.4; IRAS 10190+1322, Teng et al. 2010; NGC 3314, Hudaverdi et al. 2006) 
and four are excluded from our analysis sample (section 4.1). 
These AGNs with an optical classification of an \ion{H}{2} nucleus are
a class of ``optically elusive" AGNs.
Examples of optically elusive AGNs have been reported from 
infrared-selected samples. Classical examples are
the discovery of Compton-thick AGN in infrared bright starburst
galaxies (NGC 4945, Iwasawa et al. 1993; NGC 6240, Iwasawa \& Comastri 1998;
Arp 299, Della Ceca et al. 2002).
Maiolino et al. (2003) compiled a sample of non-Seyfert galaxies
selected by infrared luminosity,
infrared color, and the presence of a compact radio core.
AGNs are found in at least 6 of 13 objects in their sample
and AGN fraction becomes higher for more infrared luminous samples.
Most of such elusive AGNs are found to be Compton-thick.
Brightman and Nandra (2011a) found clear evidence for AGNs
in four \ion{H}{2}-like objects in the 12 $\mu$m selected sample.
Two of them (Arp 299 and ESO~148$-$IG002) are Compton-thick.
Our results, together with these previous findings, demonstrate
the efficacy of a combination of infrared and X-ray selection to find hidden AGNs in galaxies with significant
star formation activity.

We examined infrared color ($F_{18}/F_{90}$) for the elusive AGNs
in the 18 and 90 $\mu$m sample. The $F_{18}/F_{90}$ ratios
are in the range from 0.038 to 0.067. This color is typical for 
\ion{H}{2} nuclei and much colder than usual Seyferts (Figure \ref{fig:IRcolor}).
The cold infrared color also supports that these AGNs are hidden
behind star formation activity.
Reliable intrinsic X-ray luminosities are available for Compton-thin
objects. The logarithm of intrinsic luminosities in 2--10 keV are 41.6 and 42.3
for NGC 695 and IRAS 20551$-$4250, respectively.
These luminosities are relatively low for Seyfert nuclei, and could be easily
overwhelmed by star formation activity in optical and infrared.

\section{Summary}

We cross-correlated 18 and 90 $\mu$m sources in the {\akari} PSC
and X-ray sources in the 2XMMi-DR3 catalog, and made a sample of infrared/X-ray-selected galaxies. 
As the parent sample of X-ray sources,
we used objects located at $|b|>10^\circ$ and with EPIC-PN counts in 0.2--12 keV greater than 60 counts
(60851 unique sources). Infrared sources at $|b|>10^\circ$ with reliable flux measurements at 18 $\mu$m and/or
90 $\mu$m FQUAL=3) are used. There are 43865 and 62326 18 and 90 $\mu$m sources, respectively, 
satisfying these criteria. The matched sample combining 18, 90 $\mu$m, and X-ray sources consist of 173 objects.
Most of them are at a low redshift; the highest redshifts is 0.31 and 90\% of objects are at a redshift smaller than 0.05.

The sample was divided into various activity classes and groups of various absorption column densities
derived from X-ray spectra.
Diagnostic diagrams using X-ray hardness (HR4) and X-ray CR to infrared flux density ratios
were made using 173 objects in the matched sample of 18 $\mu$m, 90 $\mu$m, and X-ray sources.
AGNs obscured by a column density greater than $10^{23}$ {\pcm} are located in the lower right part of 
the diagrams
 HR4 versus CR$_{\rm 0.2-12}$/$F_{18}$ and
 HR4 versus CR$_{\rm 0.2-12}$/$F_{90}$. We selected objects
in the region HR4$>-0.1$ and CR$_{\rm 0.2-12}$/$F_{90} <0.1$
without published X-ray results to search for obscured AGNs.
An object not detected in the 90 $\mu$m band and having HR4$>-0.1$ and CR$_{\rm 0.2-12}$/$F_{18} <1$
was also selected as a candidate obscured AGN. 
We analyzed X-ray spectra of 48 objects in total after excluding Galactic sources,
sources in complex fields, or sources with very low X-ray counts after data screening.

X-ray spectra of 26 among the 48 objects (54\%)  show clear evidence for the presence of 
absorbed AGNs. 16 objects (33\%) show a continuum absorbed by a column density
ranging from $3\times10^{22}$ {\pcm} to $1.4\times 10^{24}$ {\pcm}.
Six objects (13\%) show a strong Fe-K emission line and a flat continuum, indicating the presence of a
Compton-thick AGN.
The spectra of four objects are explained by either Compton-thin or Compton-thick AGN.
22 objects (46\%) show no clear evidence for the presence of an obscured AGN.
These objects are either very faint in hard X-rays or hardness ratio is modest
(HR4$\approx$0).

Reliable constraints on X-ray absorption are available for 
151 among 179 objects, satisfying the conditions 
HR4$>-0.1$ and (CR$_{\rm 0.2-12}$/$F_{18} <1$ or CR$_{\rm 0.2-12}$/$F_{90} <0.1$).
113 objects show clear evidence for the presence of absorbed AGNs, resulting in the
success rate of 75\%. If only objects satisfying the 18 $\mu$m condition are used,
the detection rate of absorbed AGN becomes 92\%.

At least seven objects with an optical classification of an \ion{H}{2} nucleus 
show evidence for the presence of obscured AGNs, four of which are reported
for the first time in this paper. These ``optically elusive" AGNs have cold infrared 
color ($F_{18}/F_{90}$) typical for  \ion{H}{2} nuclei. Their optical classifications
and infrared colors are consistent with the idea that the star formation activity
overwhelms their AGN in the optical and infrared wavelengths.

\acknowledgements

The authors thank an anonymous referee for constructive comments that 
improved the clarity of the paper.
This research is based on observations 
obtained with {\it XMM-Newton}, an ESA science mission 
with instruments and contributions directly funded by ESA Member States and NASA,
and {\akari}, a JAXA project 
with the participation of ESA.
This research made use of the NASA/IPAC Extragalactic Database (NED) 
which is operated by the Jet Propulsion Laboratory, California Institute of Technology, 
under contract with the National Aeronautics and Space Administration, 
and the HyperLeda database (http://leda.univ-lyon1.fr).

Facilities: {\it XMM-Newton}, {\it AKARI}

\newpage

\clearpage

\LongTables
\begin{landscape}

\begin{deluxetable}{lllccccccccccccccc}
\tableheadfrac{0.0}
\tabletypesize{\scriptsize}
\tablecolumns{12}
\tablewidth{0pt}
\tablecaption{XMM + AKARI 18 $\mu$m and 90 $\mu$m Sample
\label{tb:18_90sample}
}
\tablehead{
\colhead{XMM name} &
\colhead{Other name} &
\colhead{Redshift} & 
\colhead{CR} &
\colhead{HR4} &
\colhead{$F_{18}$} &
\colhead{$F_{90}$} &
\colhead{CR/$F_{18}$} &
\colhead{CR/$F_{90}$} &
\colhead{Class} &
\colhead{Ref} &
\colhead{$N_{\rm H}$}  &
\colhead{X-ray} &
\colhead{Ref} \\
\colhead{} &
\colhead{} &
\colhead{} &
\colhead{(s$^{-1}$)} &
\colhead{} &
\colhead{(Jy)} &
\colhead{(Jy)} &
\colhead{} &
\colhead{} &
\colhead{} &
\colhead{} &
\colhead{(10$^{22}$ cm$^{-2}$)} &
\colhead{Class} &
\colhead{} \\
\colhead{(1)} &
\colhead{(2)} &
\colhead{(3)} &
\colhead{(4)} &
\colhead{(5)} &
\colhead{(6)} &
\colhead{(7)} & 
\colhead{(8)} &
\colhead{(9)} & 
\colhead{(10)} &
\colhead{(11)} &
\colhead{(12)} & 
\colhead{(13)} &
\colhead{(14)} 
}
\startdata
2XMM J004847.1+315724  &  NGC 262/Mrk 348  & 0.0150 &  2.389  &  0.402  &  0.593  &  0.736  &  4.031  &  3.2486  &  Sy2   & 1 &   13.1$^{+0.7}_{-1.3}$  & Thin & 1\\
2XMMi J005329.8$-$084604  &  NGC 291  &0.0190 &  0.034  &  0.522  &  0.204  &  3.135  &  0.166  &  0.0108  &  Sy2   & 2 &   $>$100$^{b}$  & Thick & 2\\
2XMM J005334.9+124136  &  Mrk 1502  & 0.0589 & 8.677  &  $-$0.512  &  0.714  &  1.964  &  12.144  &  4.4174  &  Sy1   & 1 &   \nodata$^{a}$  & Un & 1\\
2XMMi J005356.2$-$703804  &  IRAS 00521$-$7054  &0.0689  & 0.281  &  0.013  &  0.595  &  0.635  &  0.472  &  0.4431  &  Sy2   & 1 &   7.2$\pm0.3$  & Thin & 3\\
2XMM J010516.8$-$582615  &  ESO 113$-$G010  & 0.0257&  4.071  &  $-$0.416  &  0.207  &  1.460  &  19.663  &  2.7877  &  Sy1.8   & 3 &   $<$0.0072  & Un & 4\\
2XMM J011127.5$-$380500  &  NGC 424  &0.0118 &  0.411  &  0.463  &  1.293  &  1.181  &  0.318  &  0.3481  &  Sy1.8  & 4 &   199$^{+32}_{-40}$  & Thick & 5\\
2XMM J011607.1+330522  &  NGC 449/Mrk 1 & 0.0159  &  0.097  &  $-$0.101  &  0.531  &  1.395  &  0.183  &  0.0698  &  Sy2   & 1 &   $>$100$^{b}$  & Thick & 6\\
2XMM J012732.4+191044  &  Mrk 359  & 0.0174 &  5.229  &  $-$0.342  &  0.187  &  1.282  &  27.917  &  4.0794  &  Sy1   & 1 &   \nodata$^{a}$  & Un & 7\\
\enddata
\end{deluxetable}

\clearpage
\begin{deluxetable}{llclllc}
\tableheadfrac{0.0}
\tabletypesize{\scriptsize}
\tablewidth{0pc}
\tablecaption{X-ray and infrared luminosities.
\label{tb:luminosity}
}
\tablehead{
\colhead{Name} &
\colhead{$\log L_{\rm X}$} &
\colhead{Ref} &
\colhead{$\log \nu L_{\nu}$} &
\colhead{$\log \nu L_{\nu}$} & 
\colhead{Note} &
\colhead{Ref} \\
\colhead{} &
\colhead{} &
\colhead{} &
\colhead{(18 $\mu$m)} &  
\colhead{(90 $\mu$m)} &
\colhead{} &
\colhead{} \\
\colhead{(1)} &
\colhead{(2)} &
\colhead{(3)} &
\colhead{(4)} &
\colhead{(5)} &
\colhead{(6)} &
\colhead{(7)} 
}
\startdata
NGC 262/Mrk 348    &  43.149 & \nodata &  43.702  & 43.092 & \nodata & \nodata  \\ 
   NGC 291    &  41.007 & \nodata &  43.447  & 43.929 & \nodata & \nodata  \\ 
   Mrk 1502    &  43.599 & \nodata &  45.009  & 44.733 & \nodata & \nodata  \\ 
   IRAS 00521-7054    &  43.408 & \nodata &  45.074  & 44.384 & \nodata & \nodata  \\ 
   ESO 113-G010    &  42.580 & \nodata &  43.722  & 43.864 & \nodata & \nodata  \\ 
   NGC 424    &  41.398 & \nodata &  43.829  & 43.087 & \nodata & \nodata  \\ 
   NGC 449/Mrk 1   &  40.659 & \nodata &  43.705  & 43.421 & \nodata & \nodata  \\ 
   Mrk 359    &  40.542 & \nodata &  43.332  & 43.464 & \nodata & \nodata  \\
\enddata
\tablecomments{Column 1: Name; 
Column 2: Logarithm of observed X-ray luminosity in the 2--10 keV band;
Column 3: Reference for X-ray luminosity;
Column 4: Logarithm of $\nu L_\nu$ at 18 $\mu$m;
Column 5: Logarithm of $\nu L_\nu$ at 90 $\mu$m;
Column 6: Distance of nearby objects used to calculate luminosities.
Column 7: Reference for distance.
}
\tablenotetext{a}{X-ray spectra are analyzed in this paper, and no clear evidence for the presence of an obscured AGN is found.}
\tablerefs{
X-ray ---
(1) De Rosa et al. 2012;
(2) Brightman \& Nandra 2011a; 
(3) Rivers et al. 2014;
(4) Patrick et al. 2012;
(5) Corral et al. 2011;
(6) Ueda et al. 2005;
(7) Malizia et al. 2007;
(8) Levenson et al. 2006;
(9) Winter et al. 2012;
(10) Matt et al. 2004;
(11) Ponti et al. 2010; 
(12) Teng et al. 2010;
(13) This work;
(14) Harrison et al. 2003;
(15) Akylas \& Gergatopoulos 2009;
(16) Shu et al. 2011;
(17) Markowitz et al. 2007;
Distance ---
(1) Jensen et al. 2003; 
(2) Yoshii et al. 2014;
(3) Ekholm et al. 2000;
(4) Bose \& Kumar 2014;
(5) Springob et al. 2009;
(6) Cortes et al. 2008; 
(7) Wu et al. 2014;
(8) Terry et al. 2002;
(9) Dalcanton et al. 2009;
(10) Parodi et al. 2000;
(11) Jacobs et al. 2009;
(12) McQuinn et al. 2010;
(13) Takats et al. 2012;
(14) Tully et al. 2009;
(15) Benedict et al. 2009;
(16) Mouhcine et al. 2005
}
\tablenotetext{~}{(This table is available in its entirety in a machine-readable form in the online journal. A portion is shown here for guidance regarding its form and content.)}
\end{deluxetable}
\clearpage
\begin{deluxetable}{lllcccccccccccccccc}
\tableheadfrac{0.0}
\tablecolumns{12}
\tablewidth{0pt}
\tablecaption{XMM + AKARI 90 $\mu$m Sample
\label{tb:90sample}
}
\tablehead{
\colhead{XMM name} &
\colhead{Other name} &
\colhead{Redshift} &
\colhead{CR} &
\colhead{HR4} &
\colhead{$F_{90}$} &
\colhead{CR/$F_{90}$} &
\colhead{Class} &
\colhead{Ref} &
\colhead{$N_{\rm H}$}  &
\colhead{X-ray} &
\colhead{Ref} \\
\colhead{} &
\colhead{} &
\colhead{} &
\colhead{} &
\colhead{} &
\colhead{} &
\colhead{} &
\colhead{} &
\colhead{} &
\colhead{(10$^{22}$ cm$^{-2}$)} &
\colhead{Class} &
\colhead{} \\
\colhead{(1)} &
\colhead{(2)} &
\colhead{(3)} &
\colhead{(4)} &
\colhead{(5)} &
\colhead{(6)} &
\colhead{(7)} & 
\colhead{(8)} &
\colhead{(9)} & 
\colhead{(10)} &
\colhead{(11)} &
\colhead{(12)} 
}
\startdata
2XMM J001106.5$-$120626  &  NGC 17/NGC 34/Mrk 938  & 0.0196 & 0.087  &  0.255  &  17.041  &  0.005103  &  Sy2  & 1 &  47.0$^{+30.4}_{-21.2}$  & Thin & 1\\
2XMM J001110.4$-$120114  &  NGC 35  &0.0199 &  0.016  &  0.517  &  1.647  &  0.009424  &  Sy1  & 2 & \nodata & \nodata & \nodata \\
2XMM J002034.9$-$705526  &  IRAS 00182$-$7112  & 0.327 & 0.025  &  0.145  &  1.064  &  0.023476  &  Sy2/L  & 3 &  $>$100$^{a}$  & Thick & 2\\
2XMM J004128.0+252958  &  NGC 214  & 0.0151 & 0.037  &  0.224  &  3.261  &  0.011204  &  Sy2/L  & 4 &  17.3$^{+13.1}_{-8.0}$  & Thin & 1\\
2XMM J012445.0+320957  &  UGC 959/Mrk 991  & 0.0351 & 0.040  &  $-$0.088  &  1.760  &  0.022500  &  Sy/L  & 5 & \nodata & \nodata & \nodata \\
2XMMi J013357.6$-$362935  &  NGC 612  & 0.0298 & 0.136  &  0.882  &  2.604  &  0.052230  &  Sy2  & 6 &  113$^{+10}_{-8}$  & Thin & 3\\
2XMM J051109.0$-$342335  &  ESO 362$-$G008  & 0.0157 & 0.059  &  0.436  &  0.671  &  0.088454  &  Sy2  & 7 &  $>$100$^{b}$  & Thick & 4\\
2XMMi J081040.3+481233  &  2MASX J08104028+4812335  & 0.0776 & 0.007  &  0.451  &  0.614  &  0.011751  &  Sy2  & 8 &  34$^{+21}_{-13}$  & Thin &5\\
\enddata
\end{deluxetable}

\clearpage
\begin{deluxetable}{lllcccccccccccccccc}
\tableheadfrac{0.0}
\tabletypesize{\scriptsize}
\tablecolumns{12}
\tablewidth{0pt}
\tablecaption{XMM + AKARI 18$\mu$m Sample
\label{tb:18sample}
}
\tablehead{
\colhead{XMM name} &
\colhead{Other name} &
\colhead{Redshift} & 
\colhead{CR} &
\colhead{HR4} &
\colhead{$F_{18}$} &
\colhead{CR/$F_{18}$} &
\colhead{Class} &
\colhead{Ref} &
\colhead{$N_{\rm H}$}  &
\colhead{X-ray} &
\colhead{Ref} \\
\colhead{} &
\colhead{} &
\colhead{} &
\colhead{(s$^{-1}$)} &
\colhead{} &
\colhead{(Jy)} &
\colhead{} &
\colhead{} &
\colhead{} &
\colhead{(10$^{22}$ cm$^{-2}$)} &
\colhead{Class} &
\colhead{} \\
\colhead{(1)} &
\colhead{(2)} &
\colhead{(3)} &
\colhead{(4)} &
\colhead{(5)} &
\colhead{(6)} &
\colhead{(7)} & 
\colhead{(8)} &
\colhead{(9)} & 
\colhead{(10)} &
\colhead{(11)} &
\colhead{(12)} 
}
\startdata
2XMM J030030.5$-$112456  &  MCG $-$02$-$08$-$039 & 0.0299 &  0.200  &  0.517  &  0.371  &  10.785  &   Sy2  & 1 &  46.9$^{+13.1}_{-11.9}$  & Thin & 1\\
2XMMi J100149.4+284709  &  3C 234  & 0.185 & 0.231  &  0.402  &  0.274  &  14.602  &   Sy1  & 2 &  36.9$^{+1.9}_{-1.8}$  & Thin & 2\\
2XMM J113240.2+525701  &  HCG 056B/UGC 06527 NED03  & 0.0264 & 0.081  &  0.730  &  0.210  &  19.028  &   Sy2  & 3 &  76$^{+21}_{-17}$  & Thin & 3\\
2XMM J121857.5+471814  &  NGC 4258  & 0.00149 &  0.443  &  0.164  &  0.449  &  8.901  &   Sy1.9  & 4 &  14.8$^{+4.3}_{-3.3}$  & Thin & 4\\
2XMM J130527.3$-$492804  &  NGC 4945  & 0.00188 & 0.413  &  0.312  &  9.945  &  0.402  &   Sy2/L  & 5 &  251$^{+12}_{-11}$  & Thick & 4\\
2XMM J132527.6$-$430109  &  NGC 5128/Centaurus A  & 0.00183  & 11.563  &  0.238  &  13.148  &  0.304  &   Sy2  & 6 &  14.1$\pm0.7$  & Thin & 4\\
2XMM J140700.4+282714  &  Mrk 668  & 0.0766 & 0.076  &  0.165  &  0.305  &  13.097  &   Sy1  & 1 &  $>$100$^{a}$  & Thick & 5\\
2XMM J224937.0$-$191627  &  MCG $-$03$-$58$-$007  & 0.0315 & 0.198  &  0.430  &  0.629  &  6.356  &   Sy2  & 1 &  27.5$^{+7.9}_{-6.2}$  & Thin & 1\\
\enddata
\tablecomments{Column 1: XMM name;
Column 2: Alternative name;
Column 3: Redshift
Column 4: EPIC-PN count rate in the 0.2--12 keV band;
Column 5: Hardness ratio HR4;
Column 6: Flux density at 18 $\mu$m in units of Jy;
Column 7: EPIC-PN count rate in 0.2--12 keV to 18 $\mu$m flux density ratio;
Column 8: Optical classification. Sy --- Seyfert, L --- LINER, HII --- HII nucleus, Unclass --- unclassified, 
1 --- type 1, 2 --- type 2, and a fractional number denotes intermediate types;
Column 9: Reference for optical classification;
Column 10: Absorption column density in units of $10^{22}$ cm$^{-2}$;
Column 11: Classification of absorption measured with X-ray.
Un --- unabsorbed or slightly absorbed ($N_{\rm H}<10^{22}$ cm$^{-2}$), 
Thin --- Compton-thin ($N_{\rm H}=10^{22}-1.5\times10^{24}$ cm$^{-2}$), Thick --- Compton-thick ($N_{\rm H}>1.5\times10^{24}$ cm$^{-2}$).
If only lower limit of $N_{\rm H}=10^{24}$ cm$^{-2}$ is obtained, such objects are regarded as Compton-thick.;
Column 12: Reference for absorption column density. 
}
\tablenotetext{a}{Compton-thick candidates judged from a large Fe-K line EW ($>$ 700 eV) and/or flat continuum.}
\tablenotetext{b}{X-ray spectrum shows no evidence for significant absorption.}
\tablerefs{Optical classification --- 
(1) de Grijp et al. 1992;  
(2) Buttiglione et al. 2009; 
(3) Coziol et al. 2004; 
(4) Ho et al. 1997;  
(5) L\'ipari et al. 1997;  
(6) Storchi-Bergmann et al. 1997;  
(7) Kewley et al. 2001;  
(8)  Veilleux et al. 1995; 
X-ray --- 
(1)  Brightman \& Nandra 2011a; 
(2) Piconcelli et al. 2008; 
(3) Guainazzi et al. 2005b; 
(4) Fukazawa et al. 2011; 
(5) Guainazzi et al. 2004; 
(6) Jenkins et al. 2005 
}
\end{deluxetable}

\clearpage
\end{landscape}

\begin{deluxetable}{lcccccc}
\tableheadfrac{0.0}
\tablewidth{0pc}
\tablecaption{Log of XMM-Newton Observations
\label{tb:Xsample}
}
\tabletypesize{\small}
\tablehead{
Name & IR$^a$ & Type$^b$ & Redshift$^c$ & Galactic $N_{\rm H}$ & ObsID$^d$ &  Net exposure \\
	&	&	&		&	($10^{22}$ cm$^{-2}$) & & (ksec) 
}
\startdata
NGC 35 & 90 & Sb &  0.019894 & 0.0220 & 0150480501 & 9.2 \\
2XMM J004330.4$-$180107 & 90 & \nodata & \nodata & 0.0180 & 0112880601 & 9.0 \\
IRAS 00517+4556 & 90 & \nodata & \nodata & 0.107 & 0501230301 & 10.4 \\
UGC 587 & 90 & Sbc & 0.049991 & 0.0343 & 0012440101 & 23.4 \\
UGC 959 (Mrk 991) & 90 & Sa & 0.035131 & 0.0525 & 0201090401 & 13.6 \\
IRAS 01356$-$1307 & 18+90 & S0-a & 0.040211 & 0.0157 &  0502020201 & 6.8\\
2MASX~J02253645$-$0500123 & 90 & \nodata & 0.053967	& 0.0231 & 0111110301 & 17.7 \\
2MASX~J03084352+4051419 & 90 & \nodata & \nodata & 0.106 & 0112880701 & 44.7 \\
\enddata
\tablenotetext{a}{Infrared sample. ``18+90'' --- 18 + 90 $\mu$m sample, 18 --- 18 $\mu$m sample, 90 --- 90 $\mu$m sample.}
\tablenotetext{b}{Hubble type taken from the HyperLeda.}
\tablenotetext{c}{The redshift with errors are determined from Fe-K line.}
\tablenotetext{d}{Observation identification number.}
\end{deluxetable}

\clearpage
\begin{landscape}

\begin{deluxetable}{llccccccccccccccccc}
\tableheadfrac{0.0}
\tabletypesize{\scriptsize}
\tablecaption{Results of X-ray Spectral Fits
\label{tb:Xspec}
}
\tablecolumns{12}
\tablewidth{0pt}
\tablehead{
\colhead{Name}	& 
\colhead{Model$^{\rm a}$}&	
\colhead{$N_{\rm H}$} &
\colhead{$\Gamma^{\rm b}$} &
\colhead{$A_{\rm pl}$} & 
\colhead{$A_{\rm ref}$} & 
\colhead{$kT$} & 
\colhead{$A_{\rm apec}$} & 
\colhead{$C$/dof} & 
\colhead{$F_{2-10}$} & 
\colhead{$L_{2-10}$} & 
\colhead{X-ray} \\
 &
 &
\colhead{($10^{22}$ cm$^{-2}$)} &
&
&
&
\colhead{(keV)} & 
& 
&
&
&
class\\
\colhead{(1)} & 
\colhead{(2)} & 
\colhead{(3)} & 
\colhead{(4)} & 
\colhead{(5)} & 
\colhead{(6)} & 
\colhead{(7)} & 
\colhead{(8)} & 
\colhead{(9)} & 
\colhead{(10)} & 
\colhead{(11)} &
\colhead{(12)} 
}
\startdata
UGC 959/Mrk 991 & zpha(zpo) & 0.0 & 0.95 & 1.39 & \nodata & \nodata & \nodata & 136.0/134 & \nodata & \nodata & \\
				& apec+zpha(zpo) & $2.4^{+1.9}_{-1.3}$ & $1.95^{+0.95}_{-0.82}$ & 7.90 & \nodata & $0.74^{+0.15}_{-0.16}$ & 0.697 & 110.0/132 & 16.7 & 0.584 & \\ 
				& apec+zpha(zoo)* & $2.2^{+0.8}_{-1.4}$ & 1.8f & 6.25 & \nodata & $0.74^{+0.14}_{-0.16}$ & 0.675 & 110.1/133 & 17.3 & 0.584 & Thin\\
IRAS 01356$-$1307 &  apec+zpha(zpo+zpha(zpo)) & 0.0 & 1.8f & 1.69 & \nodata & 0.23 & 2.40 & 196.7/173 & \nodata & \nodata &  \\
				&						& 127 & 1.8f & 7.26\\
				& 2apec+zpha(zpo+zpha(zpo))* & 0.0 ($<0.44$) &  1.8f & 0.888 & \nodata & $0.114^{+0.036}_{-0.023}$ & 6.94 & 165.1/171 & 5.80 & 1.54 & Thin \\
				&						&105 ($>15$) 	& 1.8f  & 11.7(25.9$^{\rm c}$)		&	 & $0.88^{+0.13}_{-0.15}$ & 1.14 \\
2MASX~J03084352+4051419	& apec+pha(po) & 0.0 & $-$0.43 & 0.0840 & \nodata & 0.23 & 0.944 & 221.8/214 & \nodata & \nodata \\
\enddata
\end{deluxetable}


\clearpage

\newpage

\clearpage

\begin{deluxetable}{llccccc}
\tableheadfrac{0.0}
\tablecaption{Parameters of Fe-K Emission Line
\label{tb:Fe}
}
\tablecolumns{6}
\tablewidth{0pt}
\tablehead{
\colhead{Name} &
\colhead{Model$^{\rm a}$} &
\colhead{Line Center$^{\rm b}$} & 
\colhead{$A_{\rm Gauss}$} & 
\colhead{EW} & 
\colhead{$\Delta C$} \\
\colhead{} &
\colhead{} &
\colhead{(keV)} &
\colhead{} &
\colhead{(eV)} &
\colhead{}  \\
\colhead{(1)} & 
\colhead{(2)} & 
\colhead{(3)} & 
\colhead{(4)} & 
\colhead{(5)} & 
\colhead{(6)} 
}
\startdata
UGC 959/Mrk 991 & apec+zpha(zoo) & 6.4f & 0.0($<0.0958$) & 0 ($<430$) & 0.0 \\ 
IRAS 01356$-$1307 &  \nodata & \nodata & \nodata & \nodata & \nodata\\
2MASX~J03084352+4051419 & apec+pha(po+pha(po))$^c$ & 6.4f & 0.025 ($<0.14$) & 50 ($<250$) & 0.1\\ 
IRAS 03152$-$4427 & apec+zpha(zpo+zpha(zpo)) & 6.4f & 0.026 ($<0.28$) & $47 (<360)$ & 0.1 \\
UGC 2730 & apec+zpha(zpo) & 6.4f & 0.052 ($<016$) & 450 ($<1320$) & 1.2 \\
2MASX~J05052442$-$6734358 & zpha(zpo+zpha(zpo+zga)) & 6.4f & $2.1^{+2.0}_{-0.13}$ & $240^{+120}_{-140}$ & 8.8 \\
2MASX~J05430955$-$0829274 & zpha(zpo+zga) & 6.4f & $0.14^{+0.12}_{-0.10}$ & $500^{+480}_{-350}$ & 7.2 \\ 
ESO~205$-$IG003	 & apec+zpha(zpo+zpha(zpo+zga)) & $6.38\pm0.05$ & $0.49\pm0.03$ & $80^{+38}_{-37}$ & 12.8 \\
\enddata
\tablecomments{Column 1: Galaxy Name; 
Column 2: Spectral Model used to derive the Fe-K line parameters;
Column 3: Center energy of the Gaussian line; 
Column 4: Normalization of the Gaussian in units of $10^{-5}$ photons cm$^{-2}$ s$^{-1}$ in the line;
Column 5: Equivalent width;
Column 6: The improvement in fit compared to the model without a Gaussian line. 
}
\tablenotetext{a}{The description of model components is same as in Table 5. A photon index of 1.8 was assumed for the power-law component.
``...'' menas that X-ray flux around Fe-K line is low and that no meaningful constraint on Fe-K line is obtained.}
\tablenotetext{b}{``f'' denotes fixed parameter.}
\tablenotetext{c}{If a redshifted model is used and if the line center energy is assumed to be 6.4 keV, redshift is weakly constrained as $0.046\pm0.066$ and the improvement of the fit is $\Delta C = 5.6$.}
\tablenotetext{d}{The peak at 5.8 keV seen in the spectrum was assumed to be Fe-K line at 6.4 keV in the source rest frame and the best-fit redshift of $z=0.097$ was obtained. $\Delta C$ was calculated by assuming the redshift determined from the model containing a Gaussian line.}
\end{deluxetable}

\clearpage
\end{landscape}
\begin{deluxetable}{ccccccc}
\tabletypesize{\scriptsize}
\tablewidth{0pc}
\tablecaption{Summary of Number of Objects
\label{tb:numbers}
}
\tablehead{
\colhead{Sample}	& 
\colhead{No.} & 
\colhead{X-ray} &
\colhead{Unabsorbed} &
\colhead{Thin} & 
\colhead{Thick} &
\colhead{Absorbed} \\
\colhead{} & 
\colhead{} & 
\colhead{} & 
\colhead{} & 
\colhead{} &
\colhead{} &
\colhead{Fraction} \\
\colhead{(1)} & 
\colhead{(2)} & 
\colhead{(3)} & 
\colhead{(4)} & 
\colhead{(5)} &
\colhead{(6)} &
\colhead{(7)} 
}
\startdata
18+90  & 85 & 79 (93\%) & 6	& 44	& 29 & 92\%\\
90 & 84 & 62 (74\%) 		& 31	& 21 & 10 & 50\%\\
18 & 10 & 10 (100\%)	& 1	& 7	& 2 & 90\%\\
Total & 179 & 151 (84\%) & 38	& 72	& 41  & 75\%\\
\enddata
\tablecomments{Column 1: Infrared sample. 18+90 --- 18 + 90 $\mu$m sample, 
18 --- 18 $\mu$m sample in Table 4; 90 --- 90 $\mu$m sample in Table 3.
Column 2:  Number of objects in the sample.
Column 3:  Number of objects with X-ray measurement of absorption. Fraction of objects with available X-ray absorption data is shown in parenthesis.
Column 4:  Number of unabsorbed objects with $N_{\rm H}<10^{22}$ cm$^{-2}$.
Column 5:  Number of objects absorbed by Compton-thin matter.
Column 6:  Number of objects absorbed by Compton-thick matter.
Column 7:  The fraction of objects absorbed by Compton-thin or Compton-thick matter in the sample.
}
\end{deluxetable}

\clearpage

\end{document}